\documentclass[journal]{IEEEtran}
\usepackage{algorithm}
\usepackage{algpseudocode}
\usepackage{amsmath}
\usepackage{amssymb}
\usepackage{hyperref}
\usepackage{xcolor}
\usepackage{graphicx}
\usepackage{amsmath}
\usepackage{amssymb} 
\usepackage{xcolor}
\usepackage{bm}
\usepackage{diagbox}
\usepackage{booktabs}
\usepackage{multirow}
\usepackage{algorithm}       
\usepackage{algorithmicx}   
\usepackage{mathtools}      
\usepackage{varwidth}       
\ifCLASSINFOpdf
\else
\fi
\begin{document}
%
\title{RadarPLM: Adapting Pre-trained Language Models for Marine Radar Target Detection by Selective Fine-tuning}
%
%
%

\author{Qiying~Hu,
        Yaowen~Li,
        Shengyi~Zhang,
        Chuan~Huang,
        Yu~Liu,
        and~You~He
\thanks{Qiying~Hu, Shengyi~Zhang, Chuan~Huang, Yu Liu, You He are with the Department of Electronic Engineering, Tsinghua University, Beijing 100084, China.}
\thanks{Yaowen~Li is with the Shenzhen International Graduate School, Tsinghua University, Shenzhen 518055, China..}}


%
%

\markboth{IEEE Journal}%
{Shell \MakeLowercase{\textit{et al.}}: SLRMTD: Adapting Large Language Models for Marine Radar Target Detection with Preference-aware Token Modeling}
%



\maketitle

\begin{abstract}
Recent advances in pre-trained language models (PLMs) have demonstrated their capabilities in capturing universal knowledge, making them promising for radar signal processing applications. Nevertheless, directly fine-tuning PLMs on radar signals is both computationally expensive and prone to overfitting, particularly in low signal-to-clutter ratio (SCR) environments. To mitigate both issues, an effective fine-tuning framework for PLM-based marine radar target detection is proposed. First, we design a lightweight adaptation module, enabling computationally efficient fine-tuning while preserving the pre-trained model’s general knowledge. Second, an effective selective fine-tuning strategy is developed to selectively optimize different feature patches based on their online-evaluated learning values, guiding the model to concentrate on those generalizable feature patterns and significantly reducing model overfitting to  nosiy, anomalous, or overly simple patterns during optimization. Finally, a binary classification head is retrained based on autoencoder network to further enhance detection performance. Evaluations on real-world radar datasets highlight that the proposed RadarPLM framework considerably outperforms existing models, achieving a minimum of 6.35\% gain in average detection performance under challenging low SCR conditions when using sequence features. In particular, under small-sample training conditions, RadarPLM also achieves highly significant average performance gains over prior methods, demonstrating the effectiveness of integrating the PLM.

\end{abstract}


%
\IEEEpeerreviewmaketitle

\section{Introduction}
\IEEEPARstart{R}{adar}-based detection of small marine targets presents considerable challenges, particularly in complex maritime environments characterized by low signal-to-clutter ratio (SCR) conditions. Early studies, such as those employing constant false alarm rate (CFAR) detectors, rely heavily on accurate statistical modeling to discriminate between targets and clutter. Nevertheless, capturing the inherently non-stationary and non-Gaussian nature of sea clutter proves to be extremely difficult, and mismatches in statistical assumptions often lead to elevated false alarm rates. Furthermore, as these traditional techniques typically exploit only a single amplitude feature, their detection performance is severely compromised in low SCR scenarios.

Driven by the evolution of modern wide-band radar technologies that provide enhanced range and velocity resolution, increasing research efforts have focused on exploiting multi-domain characteristics of radar echoes. In this regard, numerous hand-engineered features have been developed~\cite{chen2013detection,shui2014tri,shi2018sea,bai2023floating,zhao2021eigenvalues}, including the phase, Doppler, and time-frequency domain characteristics, have been devised for target detection. However, these methods rely heavily on hand-engineered heuristics and often fail to capture the nonlinear structures inherent in radar signal features. 

To overcome the limitations of hand-engineered features, deep learning technologies have emerged as powerful solutions for small marine radar target detection, capable of automatically capturing complex nonlinear structures inherent in radar signal features. Some studies use recurrent neural networks (RNNs)~\cite{wan2022sequence}, convolutional neural networks (CNNs)~\cite{chen2021false,qu2022false,xu2023marine,xia2023target}, and graph neural networks (GNNs)~\cite{su2021maritime}. Among all the methods mentioned above, detectors based on sequence features and RNNs~\cite{wan2022sequence} or CNNs~\cite{xia2023target} offer an excellent trade-off between inference speed and detection performance.


 Recently, the research community has experienced a paradigm shift from lightweight, task-specific neural networks toward large-scale foundation models. Pre-trained large language models (LLMs), exemplified by ChatGPT and DeepSeek, have exhibited remarkable generalization abilities across diverse and complex tasks, including long-form text generation, mathematical reasoning, and code synthesis~\cite{chang2024survey}. Motivated by these advances, an increasing body of research has extended these large models beyond the textual domain to non-textual data modalities, giving rise to multimodal large language models. Recent advancements~\cite{zhou2023one, xu2025llm, zheng2025large} have also focused on extending pre-trained large language models to domains such as time-series analysis and wireless communication signal processing, demonstrating their potential beyond natural language tasks. These works consistently report notable performance improvements over conventional lightweight task-specific models, particularly in scenarios characterized by limited labeled data.
 Inspired by these developments, this work investigates the feasibility and effectiveness of leveraging PLMs as a powerful backbone network for marine radar signal processing, aiming to enhance model robustness and generalization under challenging data conditions.

 However, adapting PLMs to marine radar signal processing remains non-trivial, primarily due to two critical challenges: the requirement for lightweight fine-tuning and the need to mitigate model overfitting under complex environmental conditions. First, the large parameter scale of PLMs necessitates lightweight and efficient fine-tuning strategies. Without such lightweight adaptation, the extensive computational and memory overhead of PLMs fine-tuning would limit their applicability in real-world radar scenarios. Lightweight adaptation strategy requires reducing the number of trainable parameters by several orders of magnitude while preserving prior knowledge embedded in PLMs. Second, directly adapting PLMs to process radar signal features often leads to severe overfitting. This issue arises from the intrinsic characteristics of radar signal features. On the one hand, radar echoes typically suffer from low signal-to-noise ratios due to complex environmental noise and sensor limitations. Besides, in small marine target detection scenarios, weak target returns are frequently submerged by ocean clutter, producing mixed-status segments that bring noisy or anomalous patterns into the radar signal features. On the other hand, different feature patterns exhibit highly imbalanced convergence behaviors during fine-tuning: simple and regular patterns tend to converge rapidly, while more informative but complex patterns remain under-learned. Consequently, previous uniform optimization strategy may lead to severe overfitting, not only to noisy or anomalous patterns but also to overly simplistic ones. This highlights the necessity of a selective fine-tuning strategy that dynamically adjusts the training emphasis based on the learning value of each feature pattern.

 To overcome the above issues, we develop a novel framework named RadarPLM (based on GPT2, an advanced open-source PLM). The proposed framework effectively bridges the gap between PLMs and radar target detection tasks with a lightweight adaptation module and a selective training strategy. First, we extract five sequence features from radar echo signals and patch them into multiple feature tokens. Subsequently, a radar target detector is built upon GPT2 and is fine-tuned through a lightweight adaptation module to efficiently align with the specific radar detection task. To mitigate overfitting, we design a preference-aware loss function, inspired by~\cite{mindermann2022prioritized}. By introducing a selective training strategy to dynamically adjust the learning weights of different feature patches according to their online-evaluated learning values. Finally, the binary classification head is trained to further improve detection performance. 


The main contributions are summarized as follows:
\begin{enumerate}
    \item We introduce RadarPLM, an effective framework that adapts the GPT-2 model, an advanced open-source PLM to marine radar signal processing tasks. To the best of our knowledge, this is the first study to offer systematic validation that modern PLMs can be effectively utilized as strong initialization models for training and optimization on radar signal features.
    \item For fine-tuning, we develop a lightweight adaptation module and design a novel selective fine-tuning strategy, to reduce fine-tuning computational overhead and significantly mitigate the risk of model overfitting.    
    \item Comprehensive experiments on widely used marine radar datasets demonstrate that RadarPLM delivers state-of-the-art detection performance, particularly under challenging low SCR conditions and in small-sample training scenarios. Moreover, the training overhead and inference speed of RadarPLM are similar to those compact baseline models, making it practical for real-world deployment.
\end{enumerate}

Our preliminary study~\cite{hu2024marine} initially demonstrated the promising prospects of PLM for radar signal prcocessing applications. In this paper, we further expand that work by introducing a lightweight adaptation module and selective training strategy for more robust and generalizable optimization, achieving better performance.

The remainder of this paper is organized as follows. Section~\ref{RW} provides a brief overview of related work. Section~\ref{PM} details the operational framework of the proposed RadarPLM. Section~\ref{E} presents the experimental results along with in-depth analysis. Finally, Section~\ref{C} concludes the paper.

 \begin{figure*}[t]
\centering
    \includegraphics[width=1\textwidth]{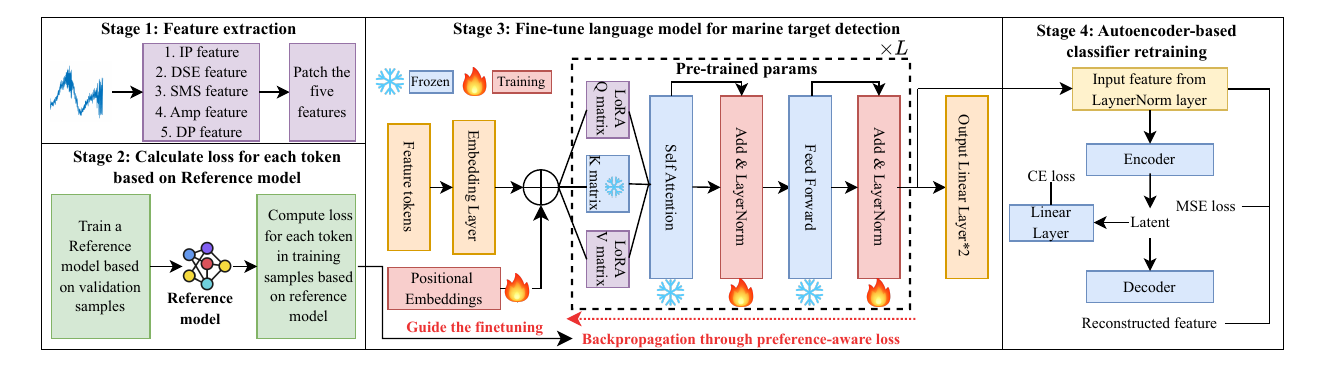}
    \caption{Overview of our RadarPLM framework.}
    \label{RadarGPT1}
\end{figure*}

\section{Related Work} \label{RW}
\subsection{Marine Radar Target Detection Based on Deep Learning}
Deep learning models with elaborately crafted architectures have demonstrated significant potentials in marine target detection. Among them, Chen \textit{et al.}~\cite{chen2021false} design a framework by combining the extracted information from both amplitude and time-frequency features to achieve target detection. Qu \textit{et al.}~\cite{qu2022false} introduce a CNN architecture augmented with asymmetric convolution layers to capture deep features of the time-frequency distribution. Xu \textit{et al.}~\cite{xu2023marine} propose a target detector in the case of limited training samples, utilizing pre-trained CNN to extract features from time-frequency spectrogram. Wan \textit{et al.}~\cite{wan2022sequence} propose a lightweight target detection framework that leverages three sequence features combining with a Bi-LSTM network. Wang \textit{et al.}~\cite{wang2022nonhomogeneous} propose a target detector based on complex value U-Net, which leverages amplitude and phase features of radar echoes for clutter suppression and subsequently detect desired targets. Su \textit{et al.}~\cite{su2024radar} introduce a graph neural network (GNN) approach for radar target detection, which constructs spatio-temporal adjacency matrices, extracts hierarchical features through convolutional graph operations, and outputs detection results via nonlinear vertex embeddings. 

In addition to traditional supervised learning, semi-supervised, unsupervised, and incremental learning approaches are proposed to further advance radar target detection. Wang \textit{et al.}~\cite{wang2023maritime} propose a self-evolving framework for marine radar target detection, which enhances detection performance in unknown environment through pseudo-labels generation, data augmentation, and model fine-tuning. Xia \textit{et al.}~\cite{xia2023target} introduce an unsupervised contrastive learning framework that learns discriminative representations between targets and clutter from unlabeled data. Wang \textit{et al.}~\cite{wang2024self} propose an incremental learning-based target detection method, which can continuously adapt the NN model in real time according to environmental changes. These methods provide compelling evidence that deep learning can significantly enhance detection performance without the need for explicit mathematical modeling. Despite their initial promise, these methods are inherently constrained by their dependence on relatively compact neural networks. Such models exhibit limited representational capacity, which hinders their ability to learn complex, high-dimensional feature representations. Consequently, they are particularly susceptible to performance degradation in the face of diverse and nonstationary environmental conditions. These challenges motivate us to first explore the use of PLMs for radar target detection. 

\subsection{Signal Processing Based on Pre-trained Language Models}
Recently, PLMs such as GPT2 and LLaMA2 have demonstrated exceptional capabilities in sequential signal processing and wireless communication tasks. For example, Zhou \textit{et al.}~\cite{zhou2023one} propose a unified time series analysis framework by fine-tuning frozen GPT2~\cite{radford2019language} and achieving SOTA performance on various datasets. Liu \textit{et al.}~\cite{liu2024llm4cp} propose a GPT2-empowered channel prediction framework to improve prediction accuracy. Sheng~\textit{et al.}~\cite{sheng2025beam} utilize the GPT2 model to develop an effective and robust beam prediction method. Zheng and Dai~\cite{zheng2025large} propose a multi-task PLM framework to satisfy the requirements of different wireless communication tasks. IOT-LLM~\cite{an2025iot} directly utilize raw signal data as input and chain of thought prompts for reasoning on the Internet of Things (IOT) tasks.  

However, none of these models are specifically devised for radar signal processing applications. In fact, recent studies~\cite{tan2024language} have raised significant concerns regarding the universal applicability and direct transferability of PLMs to various signal processing tasks. PLM-based approaches do not always deliver significant performance improvements, implying that their pre-trained parameters sometimes fail to transfer to downstream tasks. Especially, in marine radar target detection under low SCR conditions, directly fine-tuning a PLM often leads to severe overfitting issues, limiting its parameter transfer capability. To mitigate this limitation, we introduce a preference-aware loss that selectively trains on informative feature patches.
\section{Proposed Method} \label{PM} 

 \begin{figure}[h] 
\centering    
\includegraphics[width=0.5\textwidth]{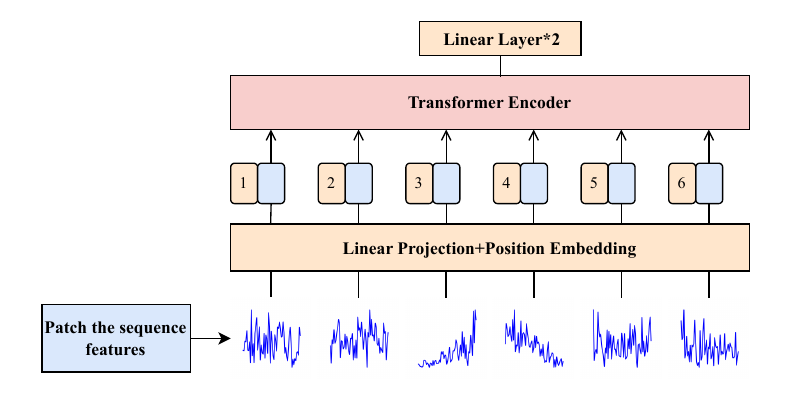}    
\caption{Overview of the reference model.} \label{RE}
\end{figure}

\begin{figure*}[htbp]
    \centering
    \begin{minipage}{0.32\textwidth}
        \centering
        \includegraphics[width=\textwidth]{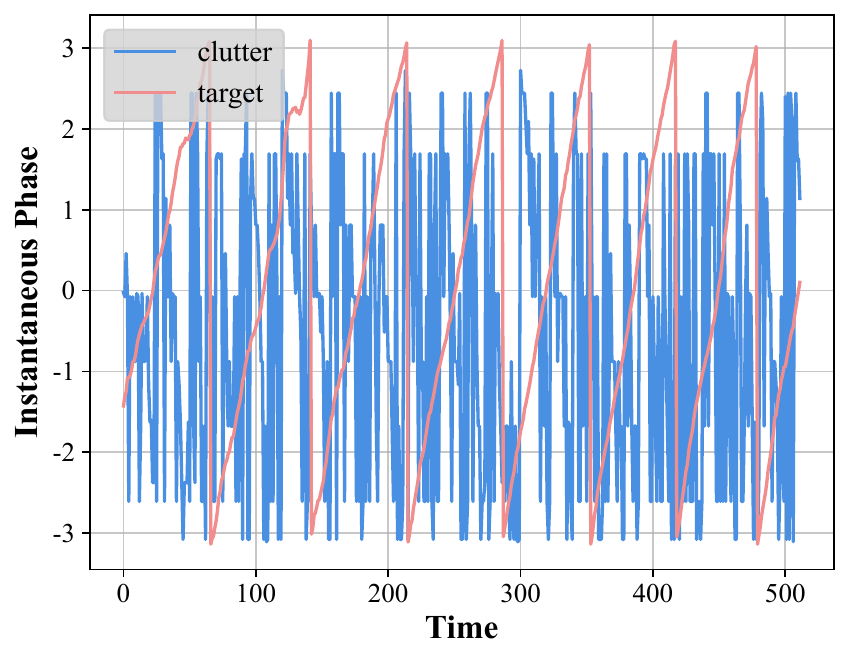}
        {\footnotesize (a) Instantaneous Phase (IP)}
        \label{fig:fig1}
    \end{minipage}%
    \hfill
    \begin{minipage}{0.32\textwidth}
        \centering
        \includegraphics[width=\textwidth]{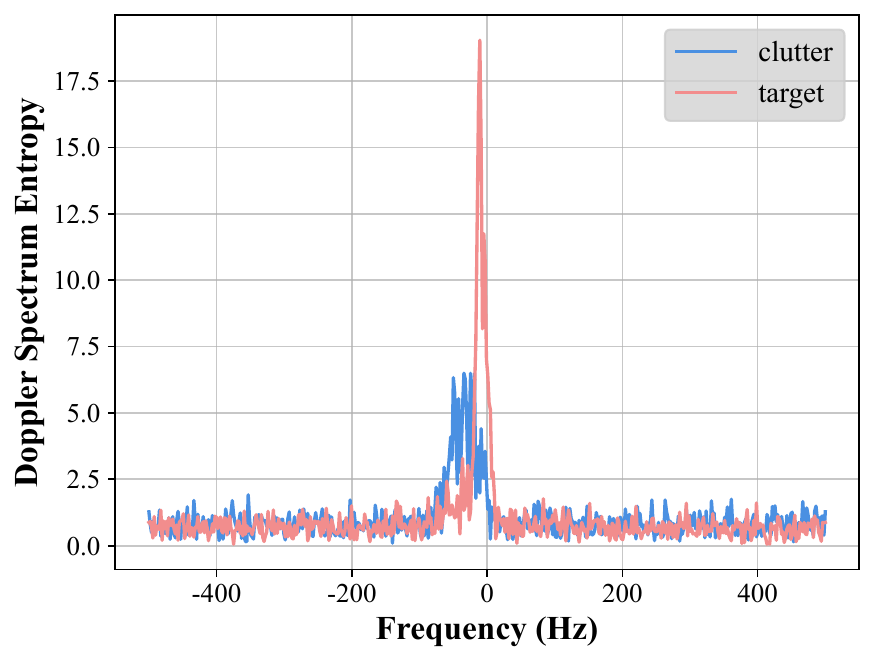}
        {\footnotesize (b) Doppler Spectrum Entropy (DSE)}
        \label{fig:fig2}
    \end{minipage}%
    \hfill
    \begin{minipage}{0.32\textwidth}
        \centering
        \includegraphics[width=\textwidth]{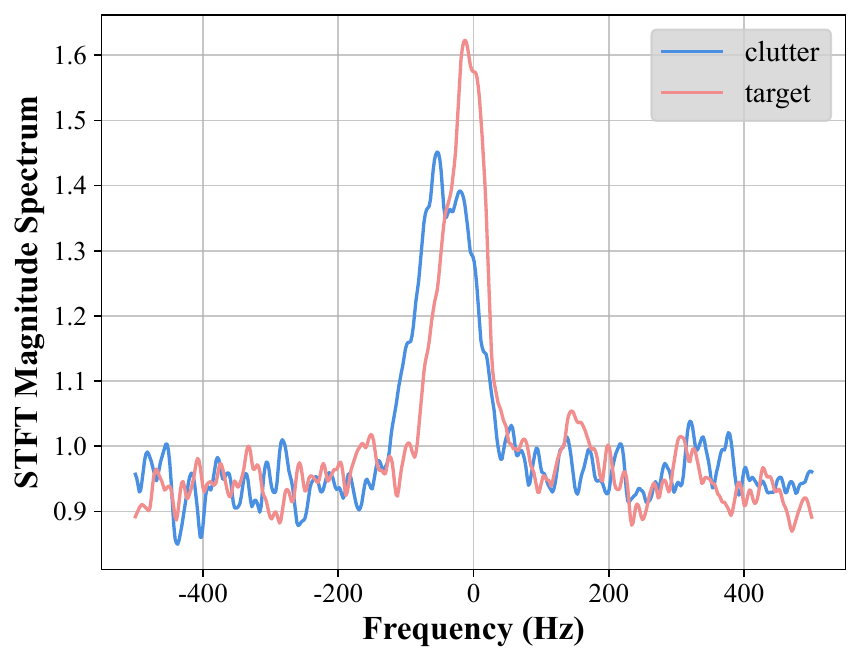}
        {\footnotesize (c) Short-Time Fourier Transform Magnitude Spectrum (SMS)}
        \label{fig:fig3}
    \end{minipage}

    \vskip\baselineskip

    \begin{minipage}{0.32\textwidth}
        \centering
        \includegraphics[width=\textwidth]{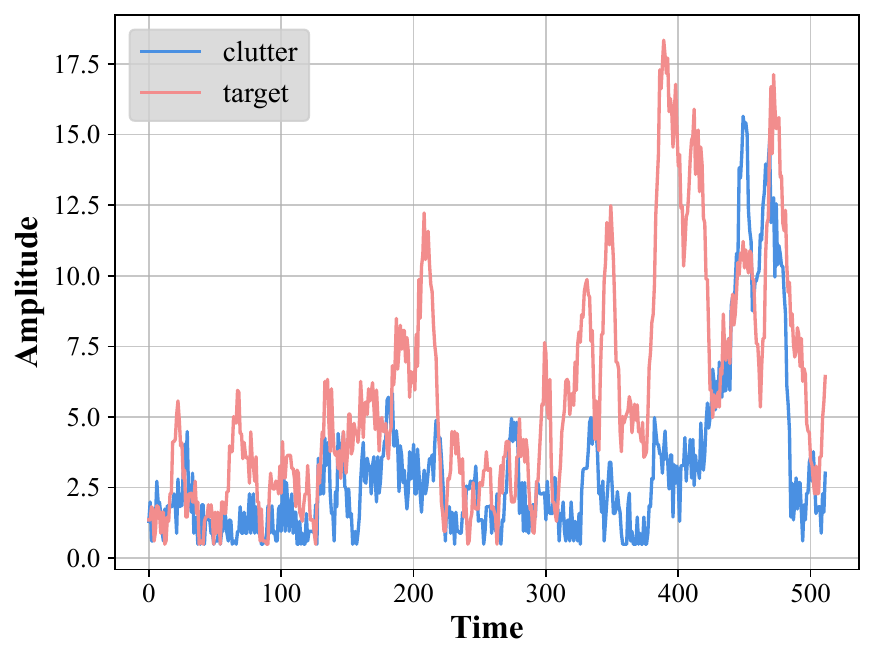}
        {\footnotesize (d) Amplitude (Amp)}
        \label{fig:fig4}
    \end{minipage}%
    \hfill
    \begin{minipage}{0.32\textwidth}
        \centering
        \includegraphics[width=\textwidth]{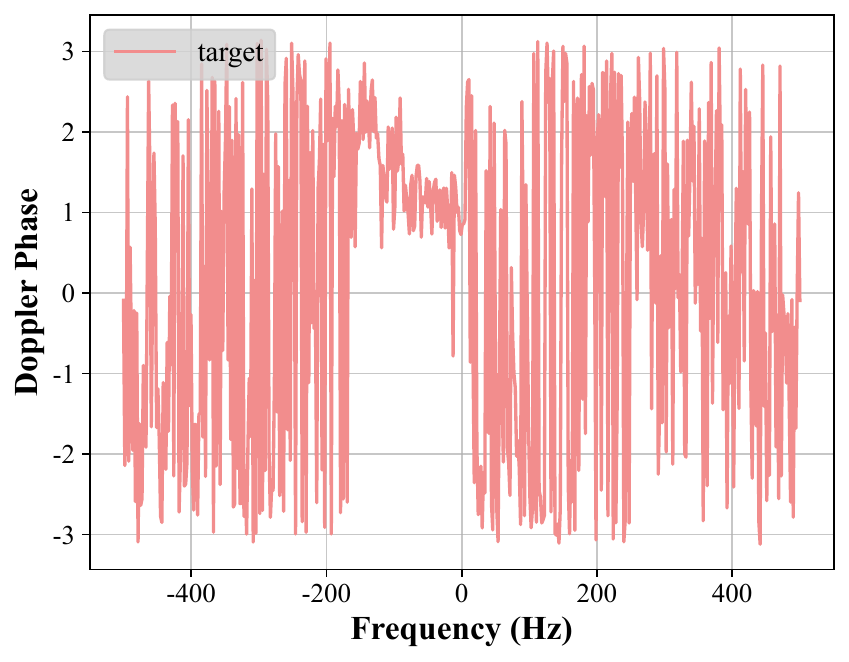}
        {\footnotesize (e) Doppler Phase (DP): Target}
        \label{fig:fig5}
    \end{minipage}%
    \hfill
    \begin{minipage}{0.32\textwidth}
        \centering
        \includegraphics[width=\textwidth]{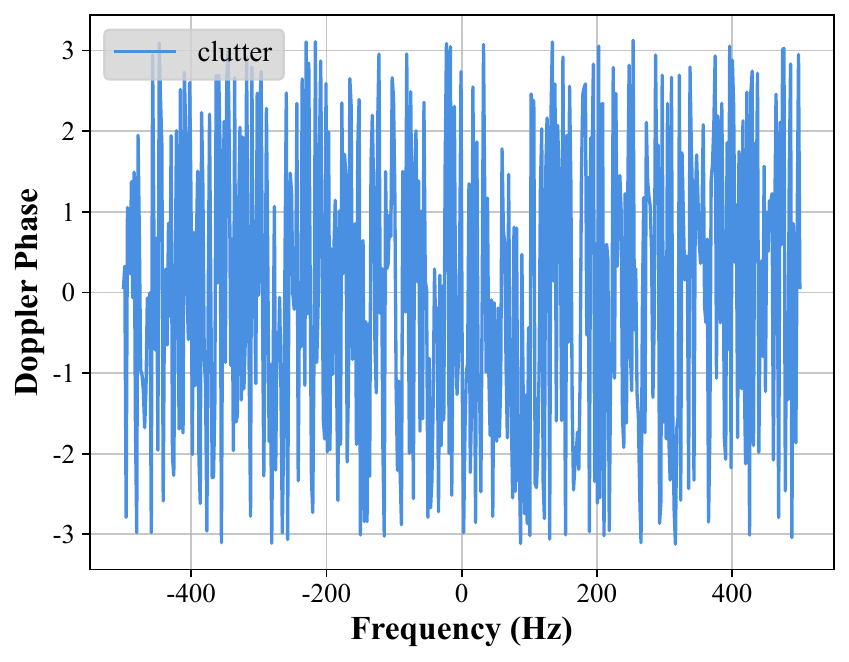}
        {\footnotesize (f) Doppler Phase (DP): Clutter}
        \label{fig:fig6}
    \end{minipage}

    \caption{IP, DSE, SMS, Amp, and DP features for target and sea clutter echo signal on IPIX dataset.}
    \label{seq_features}
\end{figure*}

\begin{figure}[h] 
\centering    
\includegraphics[width=0.4\textwidth]{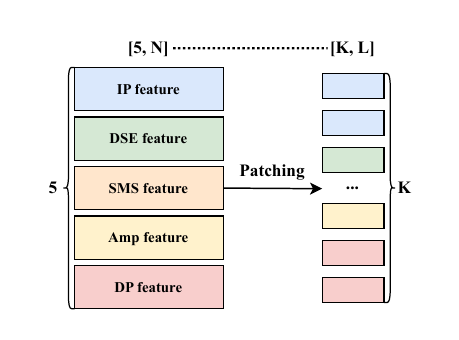}    
\caption{An illustration of the patching operation.} \label{PA}
\end{figure}

\subsection{Overview of the RadarPLM Framework}~\label{overview}

Fig.~\ref{RadarGPT1} provides an overview of RadarPLM framework, which integrates four tightly connected components: (1) sequence feature extraction and patching, (2) reference model training with token-level loss computation, (3) fine-tune LLMs for marine radar target detection, and (4) autoencoder-based binary classification head retraining.

To enable lightweight and generalizable fine-tuning of PLMs for radar target detection, the proposed RadarPLM framework integrates four tightly coupled modules into a unified optimization pipeline. First, the sequence feature extraction and patching module (Section~\ref{Module1}) derives five discriminative sequence features from the multi-domain transformation of radar echo signals. Then a patching module is devised to capture local semantic patterns while reducing computational complexity. Second, a lightweight adaptation module (Section~\ref{Module2}) is introduced to efficiently adapt the PLM backbone to the radar task, retaining its universal knowledge while minimizing parameter updates. Third, a preference-aware loss function (Section~\ref{Module3}), leveraging a lightweight reference model, enables selective optimization and mitigates overfitting. As shown in Fig.~\ref{RadarGPT1}, the PLM-based model utilizes an input embedding layer to project feature patches into the PLM’s encoder feature space. The embedded representations are processed by the initial encoder layers to extract more discriminative features. A layer normalization operation refines these features, producing the normalized representation $F^{\mathrm{LN}}$, which enhances consistency and stability. Finally, two output linear layers generate token-level detection results. For the reference model structure, a standard Transformer encoder~\cite{vaswani2017attention} processes the input feature patches to produce token-level detection outputs, as illustrated in Fig.~\ref{RE}. An autoencoder-based binary classification head (Section~\ref{Module4}) is retrained to further improve detection rate.

\subsection{Sequence Feature Extraction and Patching} \label{Module1}
When the radar transmits coherent pulses toward the sea surface, it receives a sequence of echo signals from each range resolution cell. In clutter-dominated cells, these echoes primarily consist of sea surface backscatter and thermal noise, whereas in target-present cells, additional reflections from the target are superimposed. The received radar echo sequence $x$ within a distance bin is partitioned into a series of observation vectors ${x_i}$ using the following sliding-window segmentation: 
\begin{equation} \label{Pr}
    x_i = \big[x(M \cdot (i-1) + m)\big]_{m=1}^{N}, \quad i = 1, 2, \ldots
\end{equation}
Here, $N$ denotes the window length for each observation, while $M$ defines the stride between adjacent segments. Target detection based on deep learning techniques is modeled as a binary hypothesis decision-making task:
\begin{equation}
\left\{
\begin{array}{ll}
H_0: & x(n) = c(n), \quad n = 1, 2, \ldots, N \\
H_1: & x(n) = s(n) + t(n), \quad n = 1, 2, \ldots, N
\end{array}
\right.
\end{equation}
Under the null hypothesis $H_0$, $x(n)$ contains only sea clutter $c(n)$, indicating the absence of a target. Under the alternative hypothesis $H_1$,  $x(n)$ includes both the target component $t(n)$ and the clutter $c(n)$ within the test cell. To achieve effective target-clutter discrimination, five distinct signal sequence features are derived from multiple domains, including temporal, phase, Doppler, and time-frequency domains. These sequence features include:
\begin{itemize}
    \item Instantaneous Phase (IP): The phase information of the complex-valued signal $x(n)$ is obtained as:
    \begin{equation}\label{f1}
        \pi (n)=\angle[x(n)],
    \end{equation}
    where $\angle(.)$ denotes the phase angle of the complex signal.
    \item Doppler Spectrum Entropy (DSE): First, the doppler amplitude spectrum of $x(n)$ is calculated as:
    \begin{equation}\label{f2}
        F(f_d) = \frac{1}{\sqrt{N}} \left| \sum_{n=0}^{N-1} x(n) \exp\left(-j 2\pi f_d n T_s\right) \right|,
    \end{equation}
    where $f_d(-\frac{1}{2T_s} \leq f_d \leq \frac{1}{2T_s})$ represents the Doppler frequency, and $T_s$ represents the pulse repetition interval for the radar system. Furthermore, the entropy of the Doppler spectrum is calculated as: 
    \begin{equation}\label{f3}
        \text{DSE}(F) = -\tilde{F}\left(f_{d}\right) \log \tilde{F}\left(f_{d}\right),
    \end{equation}
    where $\tilde{F}\left(f_{d}\right)=\frac{F(f_d)}{ {\textstyle \sum_{f_d}^{}}F(f_d) }$.
    \item Short-Time Fourier Transform Magnitude Spectrum (SMS): Due to the non-stationary characteristics of the radar echo signal, short-time fourier transform (STFT)  provides a more comprehensive understanding compared to Doppler transform, which is calculated as:
    \begin{equation}
        S(k,m)=\sum_{n=-\infty }^{\infty } x(n)w(n-m)e^{-j2\pi \frac{kn}{\Omega }},
    \end{equation}
    where $\Omega$ is the number of frequency bins, $w(.)$ is the hamming window function, $m$ represents the time index, and $k$ is the frequency index. To make the features suitable for input into a sequence neural network, we compute the time-averaged magnitude of $S(k,m)$ and convert it into a decibel (dB) scale. Specifically, for each frequency index $k$, the following computation is performed:
    \begin{equation}\label{f4}
        \text{SMS}(k)=10\log_{10}(\sum_{m=0}^{N-1} |S(k,m)|).
    \end{equation}
    \item Amplitude (Amp): The instantaneous amplitude is computed by calculating the magnitude of the complex radar echo signal:
    \begin{equation}
        A(n)=|x(n)|.
    \end{equation}
    \item Doppler Phase (DP): The Doppler phase is extracted from the Doppler transform of echo signal:
    \begin{equation}\label{f5}
        \text{DP}(f_d)=\angle(\sum_{n=0}^{N-1} x(n) \exp\left(-j 2\pi f_d n T_s\right)).
    \end{equation}  
\end{itemize}

As illustrated in Fig.~\ref{seq_features}, the five extracted sequence features exhibit pronounced discriminability between target and clutter samples. To better structure the input for downstream processing, the extracted features are partitioned into non-overlapping patches. This patching approach offers several key advantages: it helps retain local semantic information, reduces the computational and memory burden of attention mechanisms. For a mini-batch of $B$ samples, the input feature matrix is constructed by concatenating the IP feature $F_{\mathrm{IP}}$, DSE feature $F_{\mathrm{DSE}}$, SMS feature $F_{\mathrm{SMS}}$, amplitude feature $F_{\mathrm{Amp}}$, and DP features $F_{\mathrm{DP}}$.
\begin{equation}
    F = \text{Concat}(F_{\mathrm{IP}}, F_{\mathrm{DSE}}, F_{\mathrm{SMS}}, F_{\mathrm{Amp}}, F_\mathrm{DP}).
\end{equation}

Each feature in ${F}$ is subsequently partitioned into non-overlapping patches of length $L$. If the final patch is shorter than $L$, zero padding is applied to ensure uniform length. The resulting $K$ patches are concatenated to form the input tensor ${F}^{\mathrm{P}} \in \mathbb{R}^{B \times K \times L}$, where $K = 5\left\lceil \frac{N}{L} \right\rceil$, as illustrated in Fig.~\ref{PA}.

\subsection{Lightweight Fine-Tuning Module for the PLM Backbone} \label{Module2}
Adapting large-scale PLMs to specialized downstream applications like radar target detection presents significant challenges, necessitating a lightweight fine-tuning module. This necessity stems from two primary considerations. First, the immense parameter scale of PLMs renders full fine-tuning computationally expensive and often impractical for resource-constrained environments. Second, acquiring high-quality labeled data for marine radar target detection remains a significant challenge, often leading to model overfitting and the catastrophic forgetting of valuable knowledge embedded in PLM when performing full fine-tuning.

Drawing inspiration from recent work~\cite{zhou2023one, hu2022lora}, we propose a lightweight fine-tuning module designed to balance this trade-off between adaptability and efficiency. Our approach is guided by two hypotheses, consistent with~\cite{zhou2023one,hu2022lora}: (1) achieving strong performance does not require updating all network parameters; instead, updating a small subset, such as bias terms or specific layers, while keeping the majority frozen is sufficient. (2) The weight updates during adaptation possess a low "intrinsic rank." Consequently, instead of modifying the entire weight matrix, we can capture task-specific knowledge within a much smaller, low-rank subspace. Our module implements these insights by jointly updating the layer normalization parameters and incorporating a Low-Rank Adaptation (LoRA) component.

\begin{figure}[h] 
\centering    
\includegraphics[width=0.4\textwidth]{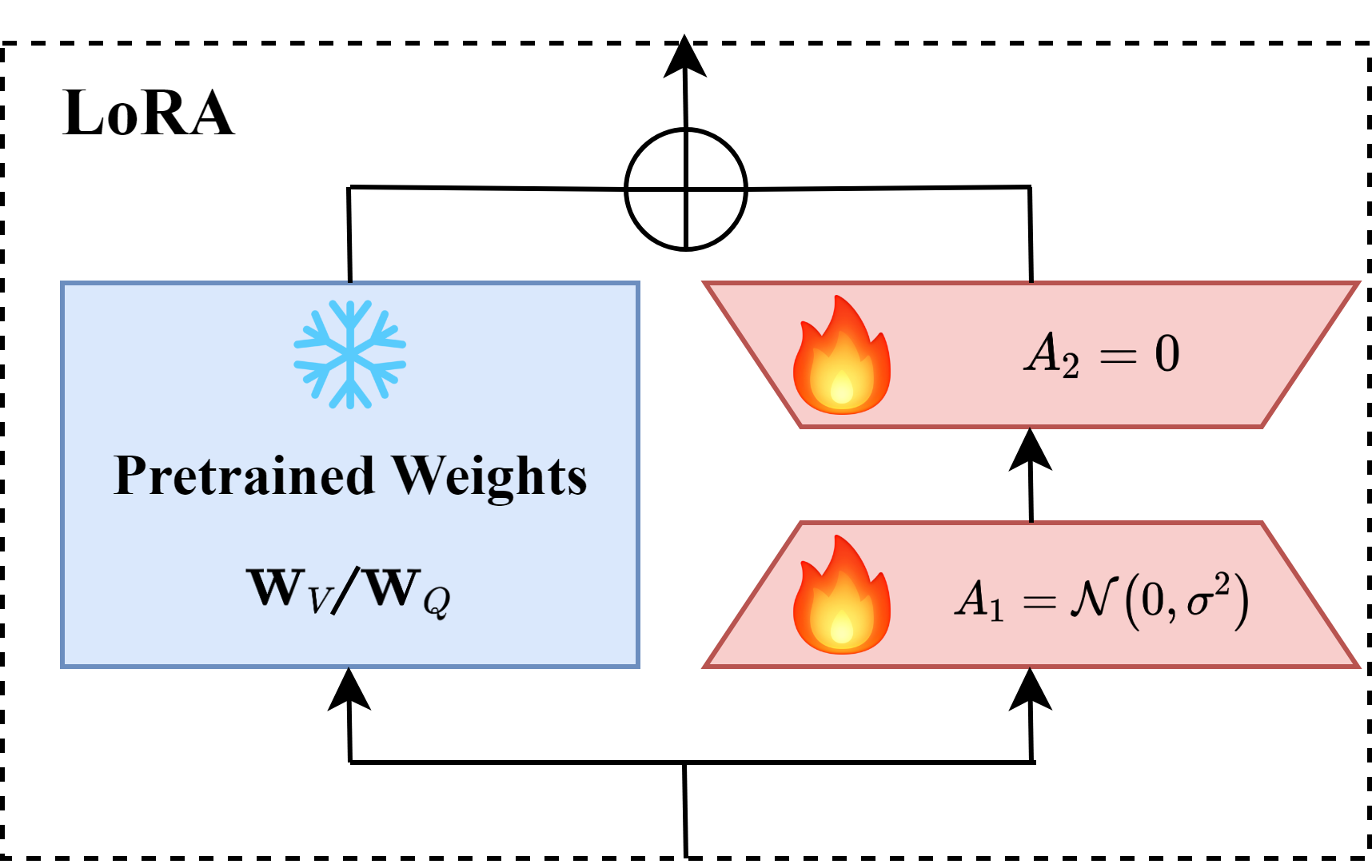} 
\centering 
\caption{Illustration of the LoRA fine-tuning process.} \label{lora}
\end{figure}

Concretely, we apply LoRA to the query and value projection matrices, $\mathbf{W}_Q$ and $\mathbf{W}_V$, within each attention block of the PLM backbone, as illustrated in Fig.~\ref{lora}. For a given frozen weight matrix $\mathbf{W}$, LoRA introduces $A_1$ and $A_2$ (low-rank trainable matrices), and modifies the transformation as follows:
\begin{equation}
\mathbf{W} \leftarrow \mathbf{W} + A_1 A_2^{T},
\end{equation}
where $\mathbf{W} \in \mathbb{R}^{l_{1} \times l_{2}}$, $A_1 \in \mathbb{R}^{l_{1} \times d}$, $A_2 \in \mathbb{R}^{l_{2} \times d}$, and $d \ll \min(l_{1}, l_{2})$. The low-rank decomposition effectively projects the task-specific adaptation into a compact subspace, greatly reducing the parameter footprint while retaining original pre-trained knowledge. The matrices are initialized as $A_1 \sim \mathcal{N}(0, \sigma^{2})$ and $A_2 = 0$, ensuring stable convergence and preventing interference with pre-trained weights during the early training phase. During fine-tuning, only $A_1$ and $A_2$ are updated, while $\mathbf{W}$ remains fixed, resulting in a trainable parameter count of $rl_1+rl_2$, which is significantly less than that of the full weight matrix, $l_1l_2$. The reduced parameter size thus makes the fine-tuning much more efficient than full fine-tuning.

This design offers two key advantages. First, this module mitigates the computationally expensive limitation of full fine-tuning by updating only a very small subset of the network parameters and compact low-rank matrices, thereby reducing trainable parameters by several orders of magnitude. Second, this module alleviates catastrophic forgetting of pre-trained knowledge during adaptation, as the newly acquired information is encoded within a subspace rather than the original representation. Consequently, this module enables the finetuned model to retain its pre-trained general-purpose knowledge while efficiently adapting to downstream radar tasks.

\subsection{Preference-Aware Loss Function} \label{Module3}

\subsubsection{Methodology}
To address the prevalent issue of overfitting in PLM-based model fine-tuning, we introduce a preference-aware loss function to enable selective training, a novel strategy designed to guide the model towards learning generalizable feature patterns. During fine-tuning, feature patches dominated by clutter-induced noise or anomalies can mislead the model and hinder effective learning. Conventional optimization methods, which treat all feature patterns indiscriminately with a standard cross-entropy loss, are prone to memorizing these non-generalizable, instance-specific characteristics, leading to severe overfitting.

In contrast, our proposed selective training strategy follows a "preference-aware" philosophy. The core idea is to evaluate the learning value of each feature patch during training and use this evaluation to reweight its contribution to the final loss. This strategy guides the model to prioritize feature patterns that provide greater learning values while minimizing the impact of noisy or anomalous patterns. By preventing the model from overfitting to these non-generalizable patterns, this strategy enhances model robustness and mitigates overfitting.

Specifically, for each feature token $F_{b, k}^{\mathrm{fi}}$ within a training batch, we compute two distinct loss values:
\begin{itemize}
    \item \textbf{Reference Loss $\mathcal{L}_{\theta_{{r}},b,k}(F^{\mathrm{fi}}_{b,k})$}: This loss is computed using a lightweight reference model $\theta_r$, which has been trained on the validation set. It serves as a stable proxy for evaluating the feature patch's inherent difficulty or noise level:
    \begin{equation} \label{r_l1}
    \mathcal{L}_{\theta_{{r}},b,k}(F^{\mathrm{fi}}_{b,k})=-\sum_{i}^{}y_{b,k}^i\log(\hat{y}_{b,k}^{i,\theta_r}),
\end{equation}
where $\hat{y}_{b,k}^{i,\theta_r}$ is the prediction from lightweight reference model $\theta_r$.
    \item \textbf{Target Loss $\mathcal{L}_{\theta_{{t}},b,k}(F^{\mathrm{fi}}_{b,k})$}: his loss is calculated using the predictions from the current PLM-based model $\theta_t$ being fine-tuned, which reflects how well the model currently fits the feature patch:
    \begin{equation} \label{t_l1}
    \mathcal{L}_{\theta_{{t}},b,k}(F^{\mathrm{fi}}_{b,k})=-\sum_{i}^{}y_{b,k}^i\log(\hat{y}_{b,k}^{i,\theta_t}),
\end{equation}
where $\hat{y}_{b,k}^{i,\theta_t}$ represents the prediction from the current PLM-based model $\theta_t$.
\end{itemize}

The key to our method is computing an evaluated learning value $s_{b,k}$ based on the discrepancy between the target and reference losses:
\begin{equation} \label{l1}
    s_{b,k}=\mathrm{ReLU}(\mathcal{L}_{\theta_{{t}},b,k}(F^{\mathrm{fi}}_{b,k})-\alpha\mathcal{L}_{\theta_{{r}},b,k}(F^{\mathrm{fi}}_{b,k})),
\end{equation}
where $\alpha$ is a scaling coefficient. To ensure that only tokens with positive excess loss are fine-tuned, we apply a $\mathrm{ReLU}(\cdot)$ operator, defined as $\mathrm{ReLU}(x)=\max(0,x)$.

This loss difference naturally leads to a selective training scheme with desirable properties:
\begin{itemize}
    \item \textbf{High Importance (Unlearned):} If $\mathcal{L}_{\theta_{{t}}}$ is much larger than $\alpha\mathcal{L}_{\theta_{{r}}}$, the feature token is identified as containing valuable, yet unlearned patterns.
    \item \textbf{Low Importance (Noisy/Anomalous patterns):} If $\alpha\mathcal{L}_{\theta_{{r}}}$ is large (indicating feature token with inherent noisy/anomalous patterns), it effectively dampens $s_{b,k}$, even if $\mathcal{L}_{\theta_{{t}}}$ is large, preventing model overfitting.
    \item \textbf{Low Importance (Well Learned):} If $\mathcal{L}_{\theta_{{t}}}$ is already small, the feature token is considered well learned, and $s_{b,k}$ will also be small, preventing redundant updates to easy patterns.
\end{itemize}

The final training loss is a weighted sum over all feature tokens across the mini-batch, where each token's contribution is scaled by a token-level importance score:
\begin{equation} \label{l2}
    \mathcal{L}_{\mathrm{final}}=\frac{1}{B* K}\sum_{b=1}^{B}\sum_{k=1}^{K} s_{b,k}\cdot \mathcal{L}_{\theta_{{t}},b,k}(F^{\mathrm{fi}}_{b,k}).
\end{equation}
where $B$ and $K$ denote the batch size and the number of feature tokens, respectively.

\subsubsection{Theoretical Insight of the Effectiveness of Feature Token Importance}
\
\par
\textbf{A. Problem Definition:} 
Inspired by~\cite{mindermann2022prioritized}, we begin by formalizing an optimization problem of selecting useful feature patches for model fine-tuning. At each training step $t$, given a candidate set $B_t$, the objective is to select a feature patch $(F_{b,k}^{\mathrm{fi}}, y_b^k) \in B_t$ whose inclusion in the training set $D_{t}$ maximally reduces the generalization loss on the unseen holdout set $D_\mathrm{ho}$. This objective can be formulated as:
\begin{equation} \label{eq1}
    \arg \min_{(F_{b,k}^{\mathrm{fi}},y_b^k)\in B_t}-\log p\left(y^{\text {ho }} \mid F^{\text {ho }} ; D_{t} \cup(F_{b,k}^{\mathrm{fi}},y_b^k)\right) .
\end{equation}

However, directly optimizing (\ref{eq1}) is computationally prohibitive, as it requires retraining the model for each candidate patch. To enable a practical surrogate, we reformulate this objective using probabilistic decomposition.

\textbf{B. Derivation of the Surrogate Objective:} Applying Bayes’ theorem, the term in (\ref{eq1}) can be decomposed as:
\begin{align}
  &-\log p\!\bigl(y^{\mathrm{ho}}\mid
      F^{\mathrm{ho}};D_t\!\cup(F_{b,k}^{\mathrm{fi}},y_b^k)\bigr)
      \nonumber\\[2pt]
  &\quad= -\log
     \frac{p\!\bigl(y, y^{\mathrm{ho}}\mid
         F,F^{\mathrm{ho}},D_t\bigr)}
          {p\!\bigl(y\mid F;D_t\bigr)}.
     \label{eq:bayes-decomp}
\end{align}

Assuming that each label is statistically independent of all other labels in the corpus when conditioned on its corresponding feature vector, the joint probability term in the numerator can be expressed as follows:
\begin{multline}
  p\!\bigl(y,y^{\mathrm{ho}}\mid F,F^{\mathrm{ho}},D_t\bigr)\\
  = p\!\bigl(y\mid F;F^{\mathrm{ho}},y^{\mathrm{ho}},D_t\bigr)\,
    p\!\bigl(y^{\mathrm{ho}}\mid F^{\mathrm{ho}};F,D_t\bigr) \\
  = p\!\bigl(y\mid F;y^{\mathrm{ho}},F^{\mathrm{ho}},D_t\bigr)\,
    p\!\bigl(y^{\mathrm{ho}}\mid F^{\mathrm{ho}};D_t\bigr).
\label{eq:joint-factor}
\end{multline}

Substituting (\ref{eq:joint-factor}) into (\ref{eq:bayes-decomp}) and taking the logarithm yields:
\begin{equation}
\begin{aligned}
\log p\!\left(y^{\text{ho}} \mid F^{\text{ho}};\, D_t \cup (F_{b,k}^{\mathrm{fi}}, y_b^{k})\right)
\\ = \log\frac{p\left(y \mid F; y^{\text {ho }}, F^{\text {ho }}, D_{t}\right) p\left(y^{\text {ho }} \mid F^{\text {ho }}, D_{t}\right)}{p\left(y \mid F; D_{t}\right)}
\\ \;\propto\;\; \log p\left(y \mid F; y^{\mathrm{ho}}, F^{\mathrm{ho}}, D_{t}\right)-\log p\left(y \mid F; D_{t}\right)
\\ \;\propto\;\; \mathcal{L}\!\left[y_{b}^{k} \mid F_{b,k}^{\mathrm{fi}} ; D_{t}\right]
- \mathcal{L}\!\left[y_{b}^{k} \mid F_{b,k}^{\mathrm{fi}} ; D_{t}, D_{\text{ho}}\right].
\end{aligned}
\end{equation}

In summary, for a model trained on $D_t$, the process of identifying the feature patch that minimizes the loss on the holdout samples in (\ref{eq1}) can be approximated by the following objective:
\begin{equation} \label{lossdiff}
    \arg \max _{(F_{b,k}^{\mathrm{fi}},y_b^k)\in B_t} \mathcal{L}\left[y_b^k \mid F_{b,k}^{\mathrm{fi}} ; D_{t}\right]-\mathcal{L}\left[y_b^k \mid F_{b,k}^{\mathrm{fi}} ; D_{\mathrm{t}},D_\mathrm{ho}\right].
\end{equation}

 Here, $\mathcal{L}[y_b^k \mid F_{b,k}^{\mathrm{fi}}; D_t]$ denotes the training loss on the patch using the current model trained on $D_t$, whereas $\mathcal{L}\left[y_b^k \mid F_{b,k}^{\mathrm{fi}} ; D_{\mathrm{t}},D_\mathrm{ho}\right]$ represents the excepted irreducible holdout loss obtained if the model were trained on both the training set and the unseen holdout set. This objective reveals that the difference between the training loss and the irreducible holdout loss effectively quantifies the reducible component of the generalization error contributed by each feature patch. Consequently, this loss discrepancy can be rigorously interpreted as the learning value of the patch, representing the marginal improvement in performance on unseen data expected from its inclusion in the training process.

\textbf{C. Practical Implementation and Approximation:}
However, realizing the irreducible loss in practice for deep neural networks presents several significant challenges. We introduce the following strategic approximations to enable an efficient and tractable implementation for calculating the irreducible loss inspired by~\cite{mindermann2022prioritized}:
\begin{enumerate}
    \item The theoretical derivation relies on Bayesian inference and conditioning. As exact Bayesian inference is intractable in neural network, we fit the model using Adam optimizer instead.
    \item In addition, we adopt the validation set ($D_{\mathrm{val}}$) as the holdout set to serve as a statistically sound proxy for the unseen data distribution.
    \item Furthermore, since training the PLM-based model on the combined set $D_{t} \cup D_{\mathrm{val}}$ is computationally expensive, we approximate it using the loss computed by a lightweight transformer-based model (reference model) trained on $D_{\mathrm{val}}$.
\end{enumerate}

\textbf{D. Conclusion:}  
  The calculated gap between the target loss $\mathcal{L}_{\theta_{{t}},b,k}(F^{\mathrm{fi}}_{b,k})$ and the approximated irreducible loss $\mathcal{L}_{\theta_{{r}},b,k}(F^{\mathrm{fi}}_{b,k})$ (derived from the lightweight reference model) serves as a practical measure of each feature patch's learning value, which forms the token weights in (\ref{l1}). Accordingly, the loss gap is employed to adaptively reweight the training objective, guiding the model to focus on learning informative and unlearned feature tokens, while significantly reducing the impact of noisy and trivial patterns.

\subsubsection{Comparison with Existing Loss Designs}
\
\par
Most existing loss improvements for marine radar target detection address issues such as sample imbalance, hard sample learning, or false alarm control. Various solutions have been proposed, including enhanced focal loss that emphasizes learning from hard or moderately difficult samples~\cite{li2025focal}, and Neyman-Pearson criterion-inspired loss that effectively regulates the false alarm rate~\cite{qu2025np}. However, these approaches seldom address a crucial issue: mitigating model overfitting in complex real-world scenarios, particularly under low SCR conditions. In contrast, the proposed preference-aware loss incorporates a selective training mechanism that directs the model’s attention to transferable and informative patterns, thus substantially reducing the risk of overfitting. Moreover, while sample reweighting strategies are often employed to alleviate overfitting by defining a weighting function that maps the training loss to a corresponding sample weight, our method adopts a patch-level reweighting strategy motivated by the uneven distribution of informative patterns across multi-domain radar features. This design enables a more fine-grained and flexible learning process. In general, the proposed preference-aware loss marks a major and substantial advance in enhancing model generalization across diverse and low-SCR radar detection scenarios.

\subsection{Autoencoder-Based Binary Classification Head Retraining} \label{Module4}
As shown in Fig.~\ref{RadarGPT1} and Section~\ref{overview}, the PLM-based network produces token-level outputs of shape $[B, K, 2]$, which is inconvenient for final decision. Accordingly, to produce a binary decision for each input, an autoencoder-based classification head is retrained on top of the high-dimensional feature representation $F^{\mathrm{LN}}$. It comprises: (1) an encoder with convolutional and fully connected layers producing compact discriminative representation; (2) a symmetric decoder ensuring feature consistency through reconstruction; and (3) a classification head operating on the latent vector. The autoencoder is optimized by a dual-objective loss combining reconstruction and classification terms:  
\begin{equation} \label{r1}
    \mathcal{L}_{\text {recon }}=\left \| F^{\mathrm{LN}}-\hat{F}^{\mathrm{LN}} \right \| _2^2.
\end{equation}
\begin{equation} \label{ce1}
    \mathcal{L}_{\text{ce}} = -\frac{1}{B} \sum_{b=1}^{B} \sum_{i=1}^{2} y_{b}^{i} \log \hat{y}_{b}^{i}.
\end{equation}
A task-uncertainty weighting scheme~\cite{kendall2018multi} adaptively balances these objectives:
\begin{equation} \label{lt}
    \mathcal{L}_{\text {total }}=\frac{1}{2 \sigma_{\text {recon }}^{2}} \mathcal{L}_{\text {recon }}+\frac{1}{\sigma_{\text {ce }}^{2}} \mathcal{L}_{\text {ce }}+\log \sigma_{\text {recon }} \sigma_{\text {ce }},
\end{equation}
where $\sigma_{\text{recon}}$ and $\sigma_{\text{ce}}$ are learnable task uncertainties that regulate the joint optimization dynamics.  

At the inference stage, the false alarm rate is regulated by applying a detection threshold derived from the ordered softmax scores of clutter-only samples. Specifically, given a target false alarm probability $P_{fa}^{\text{d}}$, the detection threshold $\eta$ is calculated as:
\begin{equation}\label{PDtH} 
\begin{array}{l} 
\eta=O_{\mathrm{sorted}}(x), \ x = \lceil P_{fa}^{\text{d}} \times N_{\text{clutter}} \rceil, 
\end{array} 
\end{equation}
where $O_{\mathrm{sorted}}(x)$ is the $x$-th largest output and $N_{\text{clutter}}$ is the total number of test clutter samples. The overall workflow of the RadarPLM is summarized in Algorithm~\ref{alg:radargpt}.

\begin{algorithm}[t]
\caption{RadarPLM Training Workflow}
\label{alg:radargpt}
\begin{algorithmic}[1]
\Require 
  \Statex Radar echo signals $\{x(n)\}_{n=1}^N$ and Target labels $\{y_b\}_{b=1}^B$ 
  \Statex PLM parameters $\theta_{\text{PLM}}$ 
\State \textbf{Stage 1: Sequence Feature Extraction}
\begin{enumerate}
    \item Extract IP, DSE, SMS, Amp, and DP features from the observation vector, respectively, see (\ref{f1}), (\ref{f2}), (\ref{f3}),(\ref{f4}), and (\ref{f5}).
    \item Patch the five features into feature tokens, see Fig.~\ref{PA}.
\end{enumerate}

\State \textbf{Stage 2: Reference Model Training}
\begin{enumerate}
    \item Train a lightweight reference model $\theta_r$ based on validation samples and CE loss.
    \item Compute the loss for each feature token of the training samples based on the predicted output probabilities of $\theta_r$, see (\ref{r_l1}).
\end{enumerate}

\State \textbf{Stage 3: Fine-tune PLM for Target Detection}
\begin{enumerate}
    \item Compute the loss for each feature token of the training samples based on the predicted output probabilities of current model, see (\ref{t_l1}).
    \item Fine-tune PLM for target detection via lightweight adaptation module and preference-aware loss function, see (\ref{l1}) and (\ref{l2}).  
\end{enumerate}

\State \textbf{Stage 4: Binary Classification Head Retraining}
\begin{enumerate}
    \item Train autoencoder-based binary classification head based on the weighted sum of reconstruction loss and CE loss, see (\ref{r1}), (\ref{ce1}), and (\ref{lt}). 
    \item Adjust detection threshold $\eta$ for controllable FAR, see (\ref{PDtH}).
\end{enumerate}
\Ensure
  \Statex Fine-tuned RadarPLM model with controllable FAR.
\end{algorithmic}
\end{algorithm}


\section{Experiments} \label{E}
\begin{table*}[t]
    \centering
    \caption{Overview of the IPIX Dataset.}
    \label{IPIXdata}
    \begin{tabular}{c c c c c}
        \toprule
        \textbf{Label} & \textbf{Index} & \textbf{File Name} & \textbf{Target Distance Bin} & \textbf{SCR} \\
        \midrule
        1 & 17  & 19931107\_135603\_starea      & 9  & Low  \\
        2 & 25  & 19931108\_213827\_starea      & 7  & Low  \\
        3 & 26  & 19931108\_220902\_starea      & 7  & Low  \\
        4 & 54  & 19931111\_163625\_starea      & 8  & High \\
        5 & 280 & 19931118\_023604\_stareC0000  & 8  & High \\
        6 & 283 & 19931118\_035737\_stareC0000  & 10 & High \\
        7 & 311 & 19931118\_162658\_stareC0000  & 7  & High \\
        8 & 320 & 19931118\_174259\_stareC0000  & 7  & High \\
        \bottomrule
    \end{tabular}
\end{table*}

\begin{table}[]
\centering
\caption{Hyperparameters for network training.} \label{hyper}
\begin{tabular}{c|c}
\hline
Parameter       & Value                    \\ \hline
Batch size      & 64                       \\ \hline
Eval batch size & 400                      \\ \hline
Epochs for RadarPLM fine-tuning          & 500                      \\ \hline
Epochs for Autoencoder training          & 300                      \\ \hline
Optimizer       & Adam (betas=(0.9,0.999)) \\ \hline
Learning rate for PLM fine-tuning   & 0.0001                   \\ \hline
Learning rate for Autoencoder training   & 0.00001                   \\ \hline
\end{tabular}
\end{table}

In this section, we first describe the experimental setup, followed by a comprehensive evaluation of RadarPLM’s detection performance under two settings: sufficient training data and limited training data scenarios.
\begin{table*}[t]
\centering
\caption{Comparison of detection rates (\%) on sufficient training samples under different signal-to-clutter ratio (SCR) conditions.}
\label{tab:detection_comparison}
\begin{minipage}{\linewidth}
\centering
\renewcommand{\arraystretch}{1.2}
\setlength{\tabcolsep}{3pt} 
\small 

\begin{tabular}{l ccccc c ccccc}
\toprule
\multirow{2}{*}{\textbf{Method}} & \multicolumn{5}{c}{\textbf{All Data}} & & \multicolumn{5}{c}{\textbf{Low SCR Data}} \\
\cmidrule(lr){2-6} \cmidrule(lr){8-12} 

 & \textbf{HH} & \textbf{HV} & \textbf{VH} & \textbf{VV} & \textbf{AVG} & & \textbf{HH} & \textbf{HV} & \textbf{VH} & \textbf{VV} & \textbf{AVG} \\
\midrule

RNN~\cite{wan2022sequence} & 46.04 & 46.66 & 38.18 & 17.04 & 36.98 & & 9.45 & 13.00 & 11.67 & 4.47 & 9.65 \\

Bi-LSTM~\cite{wan2022sequence} & 70.39 & 73.15 & 76.48 & 60.89 & 72.48 & & 39.38 & 42.47 & 37.48 & 25.27 & 36.15 \\

GRU~\cite{cho2014learning} & 76.46 & 77.55 & 73.96 & 67.48 & 73.86 & & 48.99 & 50.46 & 49.79 & 27.33 & 44.14 \\

Transformer~\cite{vaswani2017attention} & 52.95 & 71.73 & 67.32 & 55.67 & 61.92 & & 32.31 & 41.89 & 46.43 & 15.35 & 34.00 \\

PatchTST~\cite{nietime1} & 75.23 & 75.72 & 77.49 & 66.59 & 73.76 & & 49.27 & 49.71 & 53.18 & 31.46 & 45.91 \\

ResNet18~\cite{xia2023target} & 66.24 & 80.96 & \textbf{\textcolor{blue}{82.31}} & 73.97 & 75.87 & & 50.45 & 56.18 & 53.73 & 44.54 & 51.23 \\

OFA~\cite{zhou2023one} & 74.19 & 80.38 & 79.71 & 70.36 & 76.16 & & 49.28 & 54.18 & \textbf{\textcolor{blue}{54.62}} & 26.96 & 46.26 \\

ADN18~\cite{qu2023enhanced} & \textbf{\textcolor{blue}{79.74}} & \textbf{\textcolor{blue}{83.50}} & 82.01 & \textbf{\textcolor{red}{78.02}} & \textbf{\textcolor{blue}{80.82}} & & \textbf{\textcolor{blue}{54.76}} & \textbf{\textcolor{blue}{60.04}} & 52.42 & \textbf{\textcolor{red}{46.11}} & \textbf{\textcolor{blue}{53.33}} \\

\textbf{RadarPLM} & \textbf{\textcolor{red}{82.51}} & \textbf{\textcolor{red}{84.81}} & \textbf{\textcolor{red}{84.61}} & \textbf{\textcolor{blue}{75.90}} & \textbf{\textcolor{red}{81.96}} & & \textbf{\textcolor{red}{60.03}} & \textbf{\textcolor{red}{63.19}} & \textbf{\textcolor{red}{62.27}} & \textbf{\textcolor{blue}{44.82}} & \textbf{\textcolor{red}{57.58}} \\

\bottomrule
\end{tabular}

\vspace{3pt}
\centering
\raggedright 
\textit{Note:} All results are evaluated at a false alarm rate ($P_{fa}$) of 0.005. \textbf{\textcolor{red}{Red}} indicates the best result, while \textbf{\textcolor{blue}{Blue}} denotes the second-best.
\end{minipage}
\end{table*}

\begin{figure*}[htbp]
    \centering
    \begin{minipage}{0.42\textwidth}
        \centering
        \includegraphics[width=\textwidth]{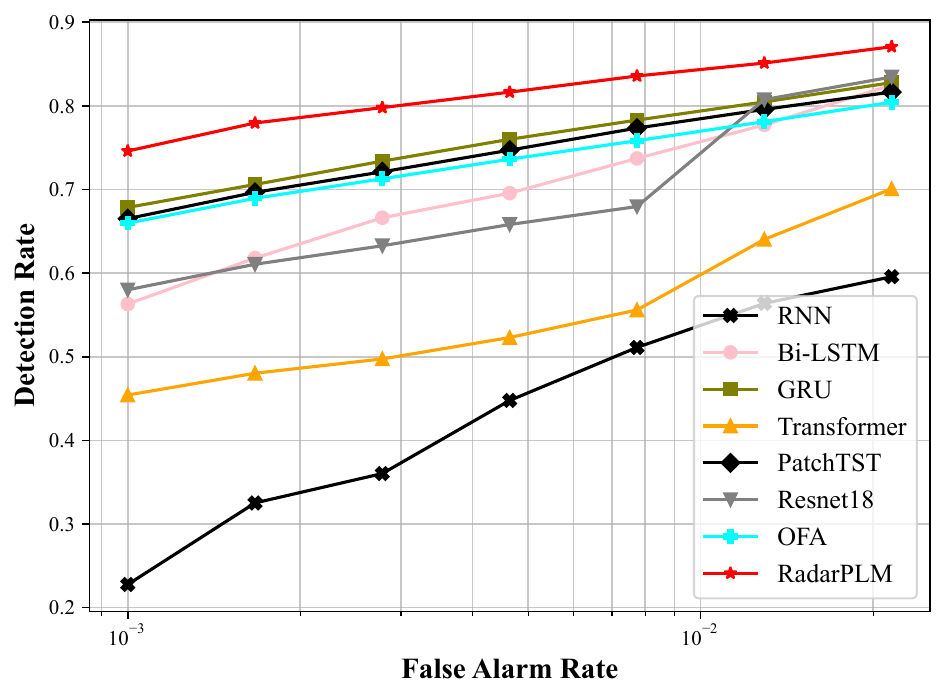}
        {\footnotesize (a). HH}
    \end{minipage}%
    \hspace{3em}
    \begin{minipage}{0.42\textwidth}
        \centering
        \includegraphics[width=\textwidth]{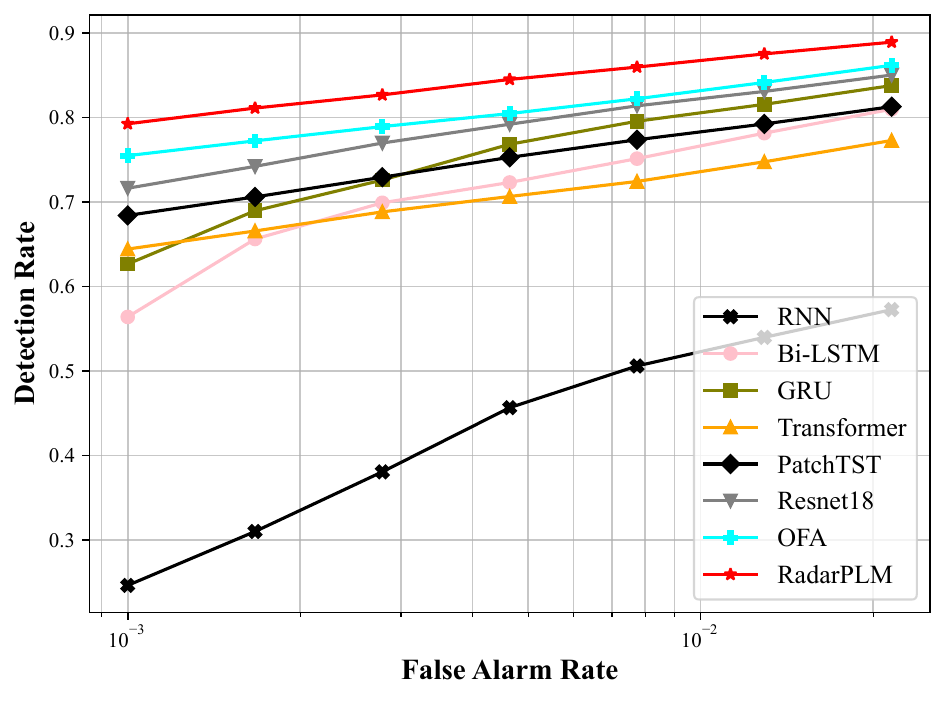}
        {\footnotesize (b). HV}
    \end{minipage}
    \vspace{\baselineskip}

    \begin{minipage}{0.42\textwidth}
        \centering
        \includegraphics[width=\textwidth]{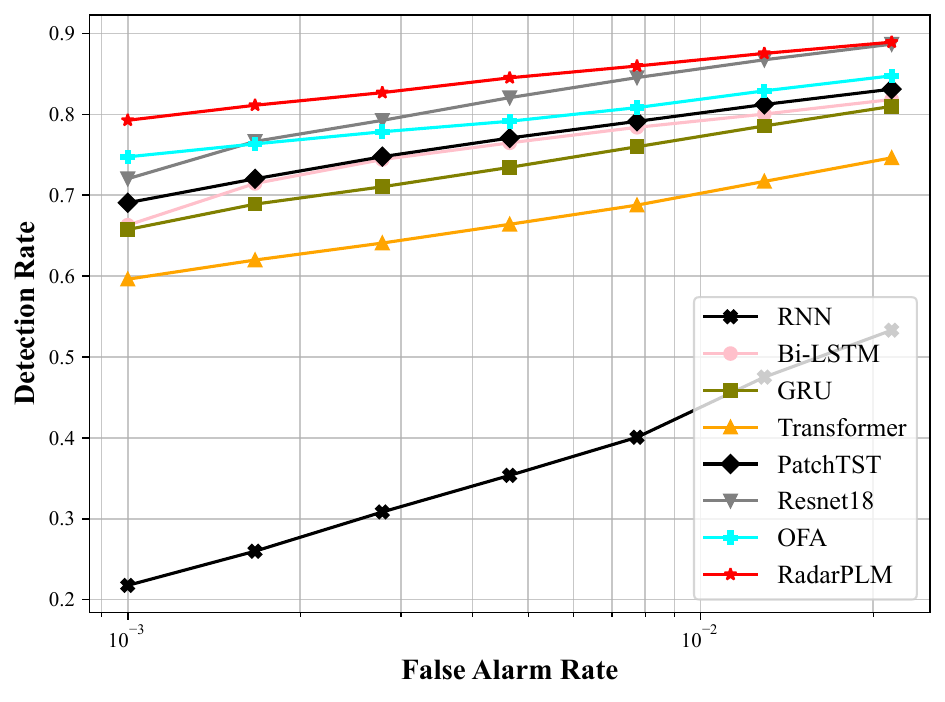}
        {\footnotesize (c). VH}
    \end{minipage}%
    \hspace{3em}
    \begin{minipage}{0.42\textwidth}
        \centering
        \includegraphics[width=\textwidth]{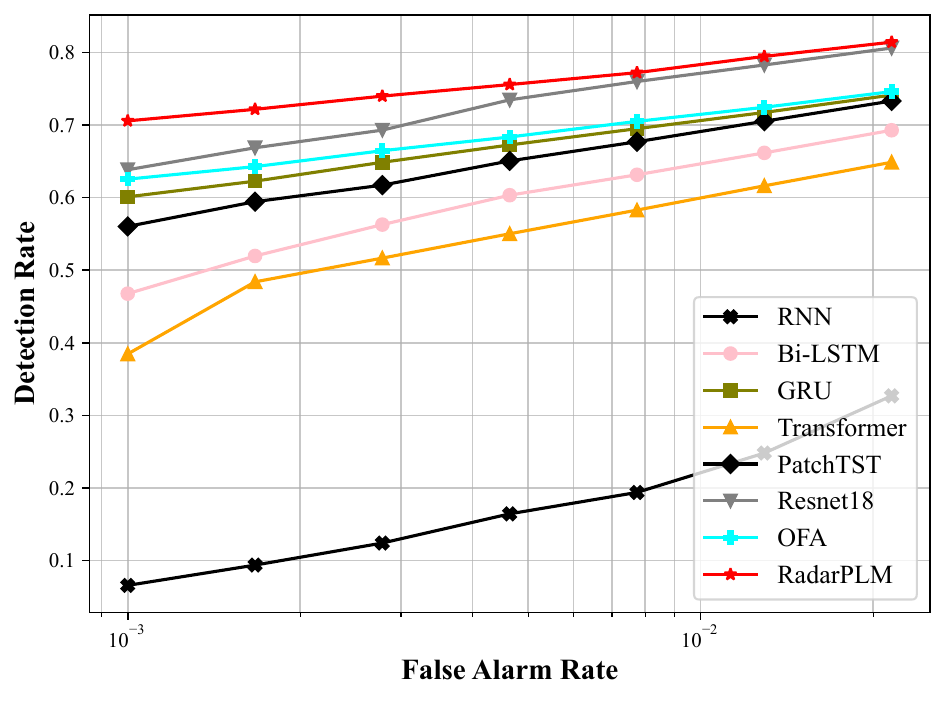}
        {\footnotesize (d). VV}
    \end{minipage}

    \caption{Comparison of model detection performance on various datasets. The accompanying ROC curves demonstrate that the RadarPLM framework achieves the best overall detection performance.}
    \label{performace_zong}
\end{figure*}
\subsection{Experimental Setup}
\subsubsection{Dataset}\label{set1}
We employ eight benchmark data sequences (Data1–Data8) from the Intelligent PIxel Processing X-band (IPIX) radar archive, summarized in Table \ref{IPIXdata}. These sequences are acquired using an IPIX radar system operating on Canada’s east coast and contain complete polarimetric returns across HH, HV, VH, and VV channels. Each set contains 14 range cells sampled at 1 kHz, with 131 072 complex echoes per cell, yielding an observation window of 0.512 s per sample. Returns from the primary range cell constitute target echoes, whereas returns from the remaining clutter-only cells represent sea clutter. To facilitate performance analysis in varying operating conditions, Data1–Data3 correspond to low signal-to-clutter ratio (SCR) scenes, while Data4–Data8 represent high-SCR scenes.

To ensure sufficient training data, we employ overlapped segmentation following the partition rule in (\ref{Pr}), with parameters set to $M=32$ for target cells and $M=128$ for clutter cells. For each dataset, the process produces more than 9,000 clutter samples, along with 4,079 target samples. The samples are divided into three groups: (1) a training set using the first 20\% observation time for both the target and clutter cells, (2) a validation set covering 20\% to 35\% of the observation time, and (3) a test set containing the remaining samples.

\subsubsection{Baselines} \label{baseline}
To validate the superiority of RadarPLM, eight deep learning-based marine radar target detection methods are implemented as baselines. Among them, RNN, Bi-LSTM, GRU, Transformer, PatchTST, and ResNet18 represent lightweight models that process sequence features; OFA uses a PLM backbone to process sequence features; and ADN18 utilizes time-frequency maps and enhanced CNNs for detection.
\begin{itemize}
    \item RNN~\cite{wan2022sequence}: Recurrent Neural Network (RNN) model temporal dependencies through recurrent connections across time steps. In our experiments, the RNN backbone consists of two recurrent layers, each equipped with six hidden units.
    \item Bi-LSTM~\cite{wan2022sequence}: Bidirectional LSTM (Bi-LSTM) extends standard LSTM by jointly modeling forward and backward temporal contexts. The implemented Bi-LSTM adopts a two-layer architecture, with six hidden units assigned to each layer.
    \item GRU~\cite{cho2014learning}: Gated Recurrent Unit (GRU) simplify the gating structure of LSTM while retaining temporal modeling capability. For evaluation, a two-layer GRU is employed, where each layer contains eight hidden units.
    \item Transformer~\cite{vaswani2017attention}: Transformer captures long-range dependencies in sequence via self-attention mechanism. The Transformer encoder is composed of two layers with a model dimension of 32 and eight attention heads per layer. Each layer includes a feed-forward network with a hidden dimension of 128.
    \item PatchTST~\cite{nietime1}: PatchTST patches sequential inputs into tokens and applies a Transformer encoder for representation learning. In our setting, the encoder has two layers, each with a model dimension of 64 and 16 attention heads, and adopts a feed-forward network with a hidden size of 128.
    \item ResNet18~\cite{xia2023target}: ResNet18 is a convolutional architecture adapted for analyzing time series by modifying input channels, leveraging residual blocks to capture temporal patterns.
    \item OFA~\cite{zhou2023one}: OFA patches input sequence features, feeds them through a frozen GPT2 model with trainable positional embeddings and LayerNorm, and uses a final linear layer for classification.
    \item ADN18~\cite{qu2023enhanced}: ADN18 enhances CNN model with asymmetric convolutional kernels and attention mechanism to adaptively capture spatial-temporal features in spectrograms derived from Short-Time Fourier Transform.  
\end{itemize}

\subsubsection{Evaluation Metrics}
In marine target detection, it is critical to maintain a high detection probability under extremely low false alarm rate constraints, since confusing sea clutter with genuine targets may lead to severe consequences, especially in military scenarios. Accordingly, detection performance is assessed at a false alarm rate of 0.005. This choice is based on the fact that the test set contains approximately 6,000 clutter samples; adopting a lower false alarm rate would yield unstable performance comparisons due to insufficient clutter samples.

\subsubsection{Implementation Details}
The experimental configuration and associated hyperparameter settings are reported in Table~\ref{hyper}. As the backbone model, the smallest GPT-2 variant is employed, using a feature embedding dimension of $F = 768$ and retaining only the first $L = 4$ encoder layers, with the patch size fixed at 48. The coefficient $\alpha$ in (\ref{l1}) is fixed at 0.9. Radar signal feature extraction is implemented in MATLAB R2016a. All training procedures are carried out on a workstation with an Intel Xeon E5-2695v3 processor, an NVIDIA RTX 3090 Ti GPU, and 64 GB of system memory.

\subsection{Average Performance Comparison} \label{Sufficient}

\begin{table*}[htbp]
\centering
\caption{Network parameters (training parameters / total parameters) and average inference time of different methods.}
\label{IN}
\renewcommand{\arraystretch}{1.1}
\setlength{\tabcolsep}{6pt}
\begin{tabular}{lccc}
\toprule
\textbf{Detector} & \textbf{Avg. inference time (s)} & \textbf{Total Params (M)} & \textbf{Trainable Params (M)} \\
\midrule
RNN~\cite{wan2022sequence}            & 5.3587 & 0.0003 & 0.0003 \\
Bi-LSTM~\cite{wan2022sequence}        & 5.6028 & 0.0016 & 0.0016 \\
GRU~\cite{cho2014learning}            & 5.7746 & 0.0008 & 0.0008 \\
Transformer~\cite{vaswani2017attention} & 6.0678 & 0.0611 & 0.0611 \\
PatchTST~\cite{nietime1}               & 5.3727 & 0.0939 & 0.0939 \\
ResNet18~\cite{xia2023target}         & 5.4434 & 3.8454 & 3.8454 \\
ADN18~\cite{qu2023enhanced}           & 160.9018 & 14.3315 & 14.3315 \\
\textbf{RadarPLM}              & \textbf{5.6561} & \textbf{69.3932} & \textbf{2.4567} \\
\bottomrule
\end{tabular}
\end{table*}

Table \ref{tab:detection_comparison} reports the detection rates achieved by RadarPLM and the eight baseline detectors introduced in Section \ref{baseline}. All evaluations are conducted on the eight IPIX datasets under a fixed false alarm rate of 0.005. Compared with the strongest sequence feature–based baseline, RadarPLM attains average detection rate gains of 7.60\%, 3.85\%, 2.30\%, and 1.93\% across the four polarization channels. The ROC curves in Fig.~\ref{performace_zong} demonstrate that RadarPLM consistently achieves the highest detection performance across the entire range of false alarm rates. We also compare RadarPLM with ADN18, which relies on high-dimensional time–frequency maps and an enhanced CNN backbone. Despite using only compact sequence features, RadarPLM still delivers a 1.14\% gain in the average detection rate.

Furthermore, we compare RadarPLM with other baseline models in terms of training complexity, total network parameters, and inference cost to evaluate its practicality for real-world deployment, with the comparison results summarized in Table~\ref{IN}. Although RadarPLM has the highest total network parameters, the number of trainable parameters is comparable to those of other models, since most weights are kept frozen during fine-tuning, thereby substantially reducing the training burden. In particular, RadarPLM achieves a similar average inference speed compared to other compact models. Specifically, it requires only 5.6561 seconds to process more than 9,000 test samples, which is 28.44 times faster than the ADN18 method. This efficiency mainly stems from the adopted patching operation, which greatly decreases the number of processed feature tokens. In addition, the inherent inference acceleration of the GPT-2 architecture further accelerate execution. Collectively, these advantages position RadarPLM as a compelling solution for practical radar system.

\subsection{Performance Comparison on Low SCR Data} \label{Low}
Owing to the inherent complexity of the marine environment, low-SCR radar echoes are frequently encountered in practical applications. In such scenarios, clutter-induced noise gives rise to noisy and anomalous feature patterns, leading to pronounced performance degradation in existing detection algorithms. To assess model robustness, we evaluate each approach on more challenging low-SCR datasets (IPIX \#17, IPIX \#25, and IPIX \#26), where the SCR typically falls below 5~dB.

As shown in the right section of Table~\ref{tab:detection_comparison}, while most baselines suffer from severe performance degradation due to strong clutter-induced noise, RadarPLM maintains a remarkable lead. It achieves an average detection rate of 57.58\% on low SCR datasets, surpassing the second-best method (ADN18) by 4.25\%. This result highlights the critical advantage of our method in learning discriminative representation from noisy and anomalous radar signal features.

\subsection{Small-sample Training Condition}
Due to the inherent sparsity of radar targets in real-world scenarios, obtaining high-quality labeled samples for radar target detection presents a substantial challenge. Considering this issue, we conducted experiments under scarce training sample conditions to assess the effectiveness of RadarPLM. Specifically, we utilized a data split ratio of 10\% (Training samples): 5\% (Validation samples): 85\% (Test samples), resulting in a deliberately limited number of training samples.

\begin{figure}[h] 
\centering    
\includegraphics[width=0.45\textwidth]{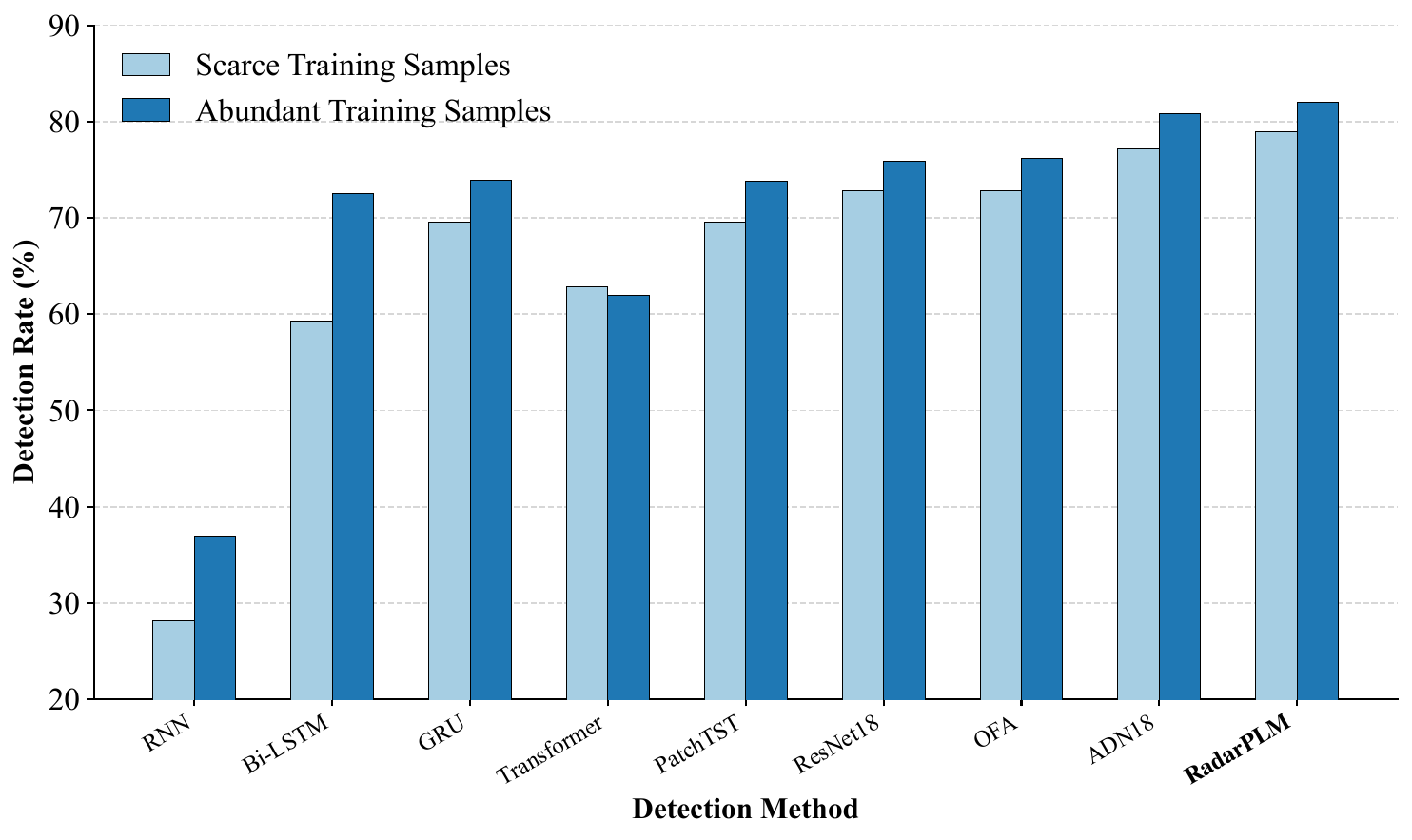}    
\caption{Performance comparison under small and abundant sample training conditions. The result shows that the RadarPLM approach experiences substantially smaller performance degradation under scarce training sample condition while maintaining the best overall detection performance.} \label{small_training}
\end{figure}

In Fig.~\ref{small_training}, we present a comparative analysis of all methods under the scarce training sample condition against the standard setting (Abundant Training Samples) described in Section~\ref{Sufficient}. The results clearly demonstrate that RadarPLM exhibits the second-lowest performance degradation among all assessed methods, highlighting its robustness to scarce data availability. Specifically, RadarPLM achieves a substantial increase of \boldsymbol{\textcolor{black}{19.66\%}} in the average detection rate over the previously proposed sequence-feature based detector~\cite{wan2022sequence}. This marked improvement can be mainly attributed to three key factors:

\begin{itemize}
\item {First}, The embedded knowledge from pre-trained parameters endows RadarPLM with a highly advantageous parameter initialization. This acts as a powerful springboard for optimization, enabling effective convergence even in data-limited regimes and drastically mitigating the data-hungry nature of conventional training paradigms.

\item {Second}, the substantial parameter scale of RadarPLM inherently grants robustness to distributional variations, enabling superior generalization compared to smaller-scale neural networks. Its extensive capacity allows it to capture complex data structures and efficiently handle variability in different radar detection scenarios.

\item {Third}, the integration of a preference-aware loss function significantly improves RadarPLM detection performance, especially in challenging scenarios characterized by low SCR.
\end{itemize}

\begin{table*}[htbp]
\centering
\caption{Ablation study on fine-tuning strategy configurations and their corresponding training time (measured over 100 training epochs).}
\label{abla1}
\renewcommand{\arraystretch}{1.1}
\setlength{\tabcolsep}{8pt}
\begin{tabular}{lcccc}
\toprule
\textbf{Model} & \textbf{LoRA} & \textbf{Layer Norm} & \textbf{Other Params} & \textbf{Training Time (hour/100 epochs)} \\
\midrule
\textbf{RadarPLM} & $\checkmark$ & $\checkmark$ &  & \textbf{0.116} \\
RadarPLM (F)    &  & $\checkmark$ & $\checkmark$ & 0.148 \\
RadarPLM (LoRA) & $\checkmark$ &  &  & 0.116 \\
RadarPLM (LN)   &  & $\checkmark$ &  & 0.112 \\
\bottomrule
\end{tabular}
\end{table*}

\begin{table}[htbp]
\centering
\caption{Results of ablation study on the backbone choice and fine-tuning strategies.}
\label{ablation}
\begin{minipage}{\linewidth}\centering 
\renewcommand{\arraystretch}{1.2}
\setlength{\tabcolsep}{3.5pt}
\begin{tabular}{cccccc}
\toprule
\textbf{Model} & \textbf{HH} & \textbf{HV} & \textbf{VH} & \textbf{VV} & \textbf{AVG} \\
\midrule
\textbf{RadarPLM} & \textbf{\textcolor{red}{82.51}} & \textbf{\textcolor{red}{84.81}} & \textbf{\textcolor{red}{84.61}} & \textbf{\textcolor{red}{75.90}} & \textbf{\textcolor{red}{81.96}} \\
\midrule
\multicolumn{6}{l}{\textbf{Backbone Choice Ablation}} \\ 
RadarPLM (0) & 77.51 & 79.38 & 80.35 & 69.97 & 76.80 \\
RadarPLM (T) & \textbf{\textcolor{blue}{78.25}} & \textbf{\textcolor{blue}{79.77}} & \textbf{\textcolor{blue}{81.16}} & \textbf{\textcolor{blue}{72.13}} & \textbf{\textcolor{blue}{77.83}} \\
RadarPLM (R) & 75.19 & 75.44 & 75.81 & 65.15 & 72.90 \\
\midrule
\multicolumn{6}{l}{\textbf{Fine-tuning Strategy Ablation}} \\ 
RadarPLM (F) & 81.69 & 83.45 & 84.41 & 74.07 & 80.91 \\
RadarPLM (LoRA) & 82.20 & \textbf{\textcolor{blue}{84.45}} & \textbf{\textcolor{blue}{84.56}} & \textbf{\textcolor{blue}{75.53}} & \textbf{\textcolor{blue}{81.69}} \\
RadarPLM (LN) & \textbf{\textcolor{blue}{82.33}} & 83.73 & 83.80 & 75.12 & 81.25 \\
\bottomrule
\end{tabular}

\vspace{5pt}

\footnotesize 
\raggedright 
\textit{Note:} RadarPLM(0)–RadarPLM(R) denote backbone ablation results, while RadarPLM(F)–RadarPLM(LN) denote fine-tuning strategy ablation results. \textbf{\textcolor{red}{Red}} indicates the best result, while \textbf{\textcolor{blue}{Blue}} denotes the second-best.
\end{minipage}
\end{table}

\begin{figure*}[htbp]
    \centering
    \begin{minipage}{0.28\textwidth}
        \centering
        \includegraphics[width=\textwidth]{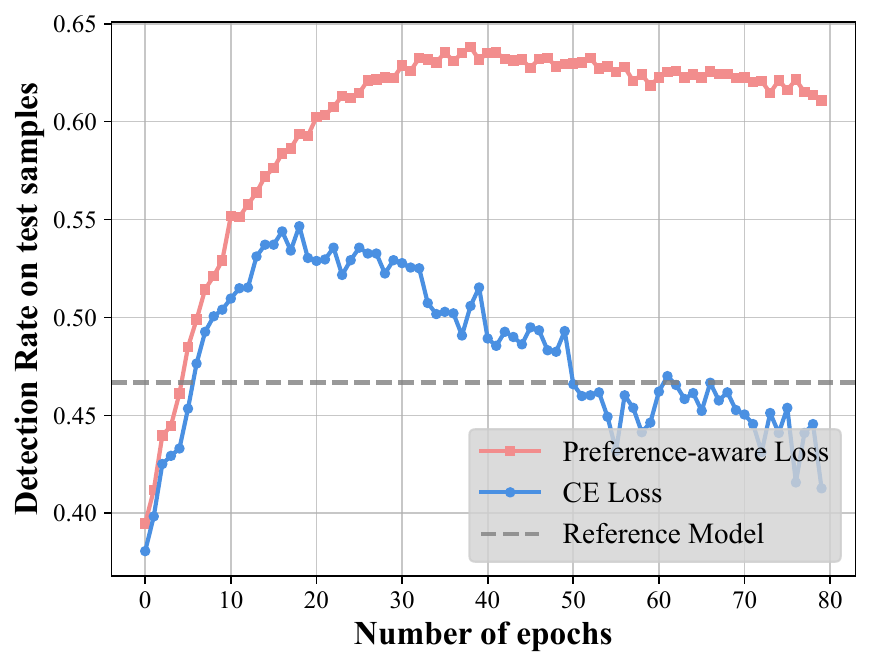}
        {(a)}
        \label{fig:fig1}
    \end{minipage}%
    \hspace{1.5em}
    \hfill
    \begin{minipage}{0.28\textwidth}
        \centering
        \includegraphics[width=\textwidth]{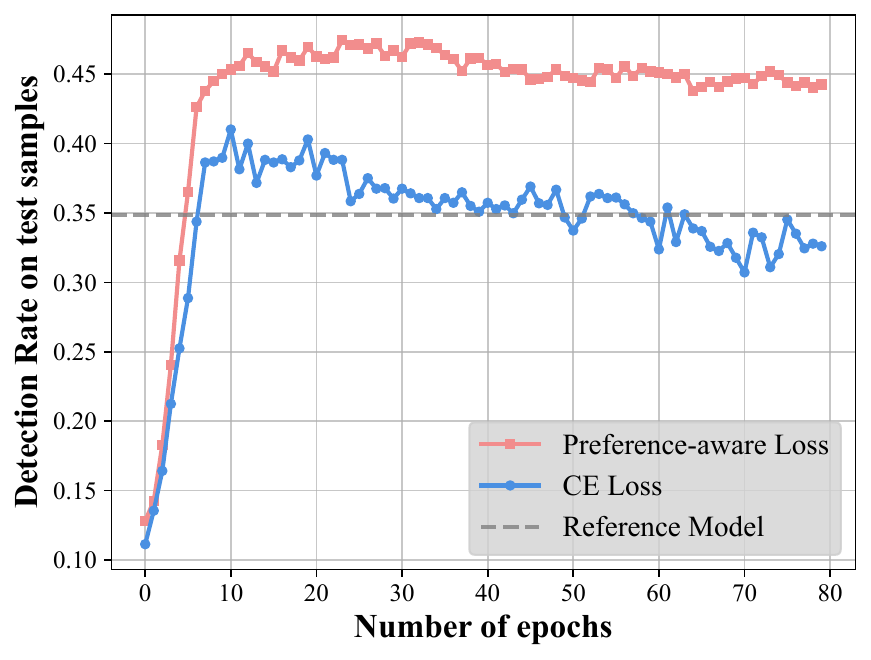}
        {(b)}
        \label{fig:fig2}
    \end{minipage}%
    \hspace{1.5em}
    \hfill
    \begin{minipage}{0.28\textwidth}
        \centering
        \includegraphics[width=\textwidth]{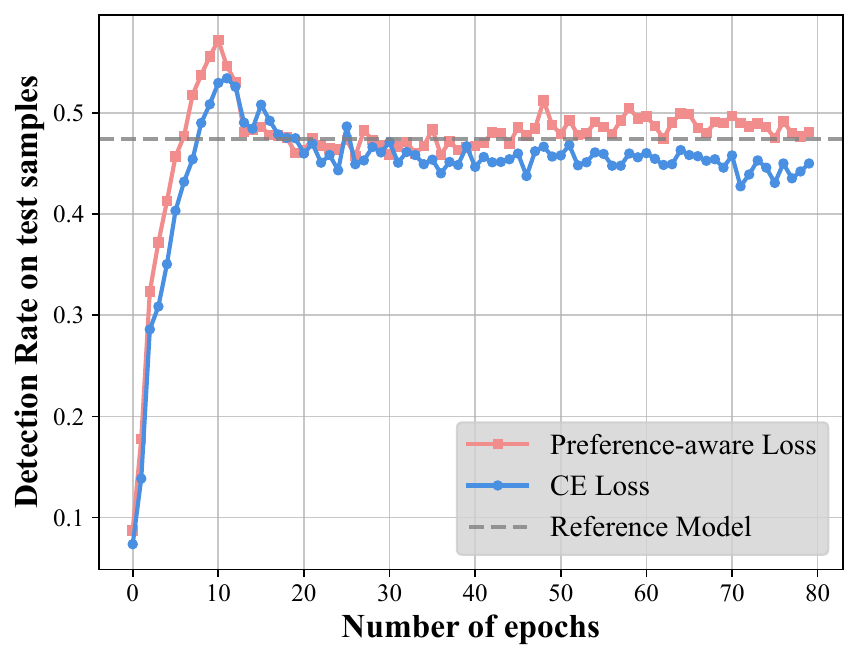}
        {(c)}
        \label{fig:fig3}
    \end{minipage}
    \hfill
    \vskip\baselineskip


    \caption{Comparison of detection rates achieved by the proposed preference-aware loss and the conventional cross-entropy (CE) loss under HH polarization across different datasets ((a) IPIX \#17, (b) IPIX \#25, and (c) IPIX \#26) over training epochs. The results demonstrate that while the CE loss exhibits detection rate degradation shortly after a few training epochs (indicating overfitting), the preference-aware loss effectively mitigates overfitting, enabling consistent performance improvement throughout more training epochs.}
    \label{features}
\end{figure*}

\begin{table}[htbp]
\centering
\caption{Results of ablation study on the loss function.}
\label{tab:ablation_loss}
\renewcommand{\arraystretch}{1.1}
\setlength{\tabcolsep}{1pt}
\begin{tabular}{ccccc}
\toprule
\textbf{Method} & \textbf{IPIX \#17} & \textbf{IPIX \#25} & \textbf{IPIX \#26} & \textbf{AVG} \\
\midrule
\textbf{RadarPLM (PA)}  & \textbf{\textcolor{red}{52.06 (+16.8)}} & \textbf{\textcolor{red}{54.87 (+5.4)}} & \textbf{\textcolor{red}{65.81 (+7.5)}} & \textbf{\textcolor{red}{57.58 (+9.9)}} \\
RadarPLM (WCE) & 43.51 \textcolor{red}{(+8.8)} & 53.95 \textcolor{red}{(+4.9)} & 64.55 \textcolor{red}{(+6.8)} & 54.84 \textcolor{red}{(+7.7)} \\
RadarPLM (CE)  & 34.73 & 48.96 & 57.80 & 47.16 \\
\midrule
\textbf{PatchTST (PA)}  & 47.48 \textbf{\textcolor{red}{(+8.1)}} & 50.29 \textbf{\textcolor{red}{(+5.9)}} & 62.42 \textbf{\textcolor{red}{(+7.0)}} & 53.40 \textbf{\textcolor{red}{(+7.0)}} \\
PatchTST (WCE) & 46.68 \textcolor{red}{(+7.8)} & 48.96 \textcolor{red}{(+5.0)} & 57.80 \textcolor{red}{(+2.9)} & 51.15 \textcolor{red}{(+5.2)} \\
PatchTST (CE)  & 38.91 & 43.92 & 54.89 & 45.91 \\
\bottomrule
\end{tabular}
\end{table}

\begin{table*}[htbp]
\centering
\caption{Performance comparison of RadarPLM under different patch sizes and GPT2 layer numbers.}
\label{tab:RadarLLM_comparison}
\begin{minipage}{\linewidth}\centering
\renewcommand{\arraystretch}{1.2} 
\setlength{\tabcolsep}{12pt}      
\begin{tabular}{ccccc}
\toprule
\textbf{Parameter} & \textbf{Configuration} & \textbf{Average DR (\%)} & \textbf{Average IT (s)} & \textbf{Total / Training NP (M)} \\
\midrule
\multirow{3}{*}{\textbf{Patch Size}} 
 & 32 & \textbf{\textcolor{blue}{79.44}} & 5.9259 & 69.39 / 2.46 \\
 & 48 & \textbf{\textcolor{red}{81.96}}  & 5.6561 & 69.39 / 2.46 \\
 & 64 & 79.09 & 5.6222 & 69.39 / 2.46 \\
\midrule
\multirow{5}{*}{\textbf{GPT2 Layers}} 
 & 0 Layer  & 76.80 & 5.3484 & 41.04 / 2.44 \\
 & 2 Layers & \textbf{\textcolor{blue}{81.29}} & 5.4812 & 55.22 / 2.45 \\
 & 4 Layers & \textbf{\textcolor{red}{81.96}}  & 5.6561 & 69.39 / 2.46 \\
 & 6 Layers & 80.42 & 5.7331 & 83.57 / 2.46 \\
 & 8 Layers & 80.01 & 5.8655 & 97.74 / 2.47 \\
\bottomrule
\end{tabular}

\vspace{5pt}

\footnotesize
\raggedright 
\textit{Note:} The results include average detection rate (DR), average inference time (IT), and total/training network parameters (NP, in millions). \textbf{\textcolor{red}{Red}} indicates the best result, while \textbf{\textcolor{blue}{Blue}} denotes the second-best.
\end{minipage}
\end{table*}

\subsection{Ablation Study}
\subsubsection{Backbone Architectures}
To examine the impact of the backbone selection, we perform an ablation study under a setting with sufficient training data, in which the PLM backbone is removed or substituted accordingly: 
\begin{enumerate}
    \item {RadarPLM (0)}: In this variant, the PLM backbone is entirely removed, while all other components of the framework are kept unchanged.
    \item {RadarPLM (T)}: In this variant, the PLM backbone is replaced with a randomly initialized transformer network.
    \item {RadarPLM (R)}: In this variant, the network architecture and fine-tuning procedure are maintained, but pre-trained parameters of PLM are replaced with random initialization.
\end{enumerate}

The results of the backbone choice ablation study are shown in Table~\ref{ablation}. The removal or exchange of PLM backbone results in notable performance degradation, indicating the necessity of PLM for high detection performance. 


\subsubsection{Fine-tuning Modules}
To assess the effectiveness of the proposed lightweight adaptation module, an ablation study under sufficient training data setting is presented in Tables~\ref{abla1} and~\ref{ablation}. In this experiment, four fine-tuning modules were evaluated: (1) RadarPLM: jointly fine-tuning the LoRA modules and Layer Normalization (LN) parameters, (2) RadarPLM (LoRA): fine-tuning only the LoRA modules, (3) RadarPLM (LN): fine-tuning only the LN parameters, and (4) RadarPLM (F): full fine-tuning of all model parameters. As shown in the results, the joint fine-tuning of LoRA and LN achieves notably higher performance, while maintaining a relatively shorter training cost (0.116 h/100 epochs) than full fine-tuning (0.148 h/100 epochs). This demonstrates that the proposed design not only enhances model adaptability, but also significantly reduces computational cost.


\subsubsection{Loss Function}
In addition, we compare the fine-tuning effects of preference-aware loss and CE loss on low SCR datasets. First, to facilitate a direct and quantitative comparison between the two loss functions, we aggregated the outputs of all feature tokens by applying a softmax normalization function followed by averaging operation. The resulting probabilities are then used for target detection. In Fig.~\ref{features}, we present the detection rates on test samples from different low-SCR datasets under HH polarization across multiple training epochs. It can be observed that preference-aware loss achieves an improvement of around \boldsymbol{\textcolor{black}{1\%-12\%}} compared to cross-entropy loss (uniformly optimized on all feature patches) in different training epochs and datasets. We also include the performance of the reference model (gray line) for comparison. The results indicate that the fine-tuned PLM-based model consistently surpasses the reference model, achieving approximately \boldsymbol{\textcolor{black}{3\%-18\%}} improvements across various datasets. Notably, despite the pronounced performance gap, leveraging the reference model to formulate the preference-aware loss still yields substantial improvements, indicating that the proposed approach effectively achieves a strong weak-to-strong generalization capability.

Although the preceding results are promising, a direct comparison between the proposed preference-aware (PA) loss and the conventional end-to-end cross-entropy (CE) loss remains lacking. To bridge this gap, we evaluated multiple model variants: RadarLLM (PA) vs. RadarLLM (CE), and PatchTST (PA) vs. PatchTST (CE). Models labeled with (PA) adopt the proposed token-level reweighting strategy based on learning value, whereas those labeled with (CE) rely on standard cross-entropy loss. For further comparison, we include a sample-level reweighting baseline, denoted as WCE~\cite{mindermann2022prioritized}, where training weights are adjusted at the sample level.

As summarized in Table~\ref{tab:ablation_loss}, the PA loss consistently yields superior performance across all datasets, especially under low SCR conditions. For example, RadarPLM (PA) achieves an average gain of \boldsymbol{\textcolor{black}{9.9\%}} over RadarPLM (CE), while PatchTST (PA) improves by \boldsymbol{\textcolor{black}{7.0\%}} over its CE-based counterpart. In particular, token-level reweighting in both PA variants outperforms WCE, highlighting the advantage of fine-grained reweighting in enhancing detection under challenging scenarios.

\subsection{Analysis of the Effect of Hyperparameters on Experimental Results}


In this section, the effect of patch size on RadarPLM performance is evaluated, and the results are provided in Table~\ref{tab:RadarLLM_comparison}. Among the evaluated settings, a patch length of 48 yields the highest detection rate, surpassing both smaller (32) and larger (64) patch configurations. This observation underscores the critical role of appropriate patch granularity in achieving optimal performance.

In addition, we analyze the effect of varying the number of GPT-2 encoder layers. The ablation results for different layer configurations are also summarized in Table~\ref{tab:RadarLLM_comparison}. Activating four GPT2 layers achieves the highest detection rate, outperforming both shallower configurations and deeper ones. These results highlight two key observations: (1) a model with insufficient layers fails to fully leverage the parameter transfer capabilities of the PLM, while excess layers suffer from overfitting due to the task-irrelevant parameters in deeper layers. (2) Although utilizing more GPT2 layers significantly increases the scale of model parameters, the impact on inference latency is minimal. Specifically, the inference speed of GPT2 (8) remains approximately 91\% that of GPT2 (0), as the inherent inference acceleration of the GPT architecture mitigates the computational overhead introduced by deeper layers.

\section{Conclusion} \label{C}
We propose a fine-tuning framework RadarPLM that adapts the GPT2, an advanced open-source pre-trained language model, for marine radar target detection. To ensure lightweight and robust adaptation, we devise a novel lightweight adaptation module, preserving the general knowledge of the GPT2 model while significantly reducing training time and fine-tuning overhead. To mitigate overfitting on low-SCR radar signal, we develop a selective training strategy  to selectively optimize different feature patches based on their online-evaluated learning values, effectively improving model robustness. Comprehensive evaluations on real-world marine radar datasets show that RadarPLM delivers consistently superior performance compared with SOTA baselines, particularly in low-SCR scenarios, with an average detection rate gain of over 6\% relative to the strongest competing methods. Notably, under small-sample training conditions, RadarPLM surpasses the existing methods by nearly 20\% at most, owing to the effective transfer of the universal knowledge in GPT-2 to the marine radar signal processing task.

\ifCLASSOPTIONcaptionsoff
\fi



%

\bibliographystyle{IEEEtran}

\bibliography{bibtex/bib/IEEEexample}

%




\end{document}